\documentclass[journal=jpccck,manuscript=article]{achemso}
\usepackage{amssymb}
\usepackage{amsmath}
\usepackage{graphicx}
\usepackage{dcolumn}
\usepackage{bm}

\title[Electron-Vibrational Interaction in Polariton
Luminescence]{Effects of Electron-Vibrational Interaction in Polariton
Luminescence: Non-Markovian Fano Resonances and Hot Luminescence}

\author{B. D. Fainberg}
\affiliation{Faculty of Sciences, Holon Institute of Technology, 52 Golomb St., Holon
5810201, Israel}
\alsoaffiliation{School of Chemistry, Tel-Aviv University, Tel-Aviv 69978, Israel}
\email{fainberg@hit.ac.il}
\author{V. Al. Osipov}
\affiliation{Faculty of Sciences, Holon Institute of Technology, 52 Golomb St., Holon
5810201, Israel}

\begin{document}

\maketitle

\begin{abstract}
{\small We have developed a non-Markovian theory of the polariton
luminescence taking the molecular vibrations into account. The calculations
were made in the polariton basis. We have shown that the frequency shift and
the polariton spectral lines broadening strongly depend on the
frequency-dependent exciton contribution to the polariton. In the
single-mode microcavity our non-Markovian theory predicts the Fano
resonances in the polariton luminescence, and also narrowing of the spectral
lines with the increase of the total number of molecules in the case of the intramolecular
nature of the low frequency vibrations. The theory enables
us to consider a non-equilibrium (hot) exciton-polariton luminescence
similar to the hot luminescence of molecules and crystals. This opens a way
for its observation in organic-based nanodevices.}
\end{abstract}

\date{\today }

\section{Introduction}

In recent years the Frenkel exciton polaritons (EPs) in organic materials
such as dye molecules\cite{Takazawa10,Fainberg17APL}, polymers and
biological nanostructures\cite{Gather16}, have attracted considerable
interest in the context of the polariton chemistry\cite%
{Ebbesen12,RibeiroPolChem18,Nitzan22ARPC} and due to their property to form
the EP Bose-Einstein condensate\cite{Plumhof14} leading to the macroscopic
coherence and superfluidity\cite{Lerario17}. The process of condensation
opens a practical possibility for manufacturing of the low threshold
polariton lasers with conventional nanosecond excitation\cite{Gather16}.

The concept of EPs suggests the presence of the strong interaction between
light and matter\cite{Agranovich03Thin_Films,Agranovich09,Hopfield58}. This
effect becomes well pronounced especially in organic dyes. The oscillator
strength of the organic dye molecules is known to be large compared to
inorganic semiconductors. This results in a large separation between the
lower (L) and upper (U) polariton branches, which can reach values\cite%
{Takazawa10,Fainberg18Advances} $\sim 0.5-1eV$.

The absorption spectrum of a large molecule manifests a progression of
spectral lines shifted with respect to each other on the frequency of a
high-energetic optically active (OA) vibration (a vibrational mode that
reorganizes after electronic excitation) $\omega _{0}\approx 1200-1500$ $%
cm^{-1}$. Polariton absorption of the electron-vibational systems was
studied in a microcavity using a model of a single high-frequency (HF) OA
vibration with a phenomenological constant decay rate (Markovian relaxation)
\cite{Rocca09Linear}; in organic dye nanofibers using a mean-field theory
for intermolecular interactions and a realistic non-Markovian model of HF
and low frequency (LF) OA molecular vibrations \cite{Fainberg19JPCC}. In a
single mode cavity the polariton absorption was investigated in the presence
of Brownian dissipation from molecular vibrations \cite{Toppari21JCP}, and
also in the Langevin approach \cite{Reitz19,Reitz20PRR}.

The luminescence of EPs attracted attention immediately after introduction
of the polariton concept. The clear physical picture of the polariton
relaxation formulated for a weak exciton-phonon interaction \cite{Toyozawa59}%
, however, is not applicable to the molecular systems, where the
electron-vibration interactions are strong. In this case, both the
interaction with radiation field and the electron-vibrational interaction
must be regarded as equally strong, which makes solution of the problem a
challenging task \cite{Toyozawa59}. Equilibrium polariton luminescence of
molecular electron-vibrational systems in cavity was calculated using
numerical solution of coupled equations of motion \cite{Lidzey08} or balance
equations \cite{Rocca09}, where the authors took a single HFOA
intramolecular vibration into account. They introduced also complex exciton
replicas frequencies with an imaginary part describing the phenomenological
constant damping rates (Markovian relaxation) \cite{Rocca09Linear}. In a
more realistic situation, however, the relaxation of molecular, exciton and
polariton systems are essentially non-Markovian \cite%
{Ebbesen15,Keeling2018NatCom} and cannot be described by means of the
constant decay rates, which can only result in the Lorentzian shape spectra.
To this end the Brownian dissipation model has been used in Ref.\cite%
{Toppari21JCP}. The authors focused on the elastic cavity emission and
restricted their consideration onto the specific case, when the polariton
spectra linewidth is the mean of the cavity and the molecular linewidth,
under the resonance condition, i.e. when the cavity eigenfrequency is in
resonance with the molecular excitation transition. However, the
luminescence is essentially an inelastic process, where the interaction of
the polariton with molecular vibrations should depend on the exciton
contribution to the polariton.

In this work we calculate the polariton luminescence, working in the
polariton basis and using the non-Markovian theory for the description of
the polariton-molecular vibration interactions. We show that the frequency
shift and the polariton spectrum broadening strongly depend on the exciton
contribution to the polariton, which is a function of frequency. Since the
proper description of the Stokes shift is very important for a correct
luminescence theory, we adopt here a realistic model, according to which
each member of the progression with respect to a HFOA molecular vibrations
of the absorption and luminescence molecular spectra is broadened due to the
presence of the LFOA vibrations $\{\omega _{\sigma }\}$. The latter makes
the main contribution to the Stokes shift in dyes \cite%
{Fai00JCCS,Fai03AMPS,Fainberg18Advances,Fainberg19JPCC}.

It is usually believed that the polariton states are formed when the
splitting of the upper and the lower polariton branches is larger than the
broadening of the molecular resonances. In contrast, our non-Markovian
theory demonstrates (i) the main characteristic features of the Fano
resonance in the polariton luminescence spectra, which appears due to the
interference of the contributions from the upper and lower polariton
branches in the regime of small splitting; and (ii) the motional narrowing of the EP
luminescence spectrum in the case of the intramolecular nature of the LFOA
vibrations with an increase in the number of molecules in the single mode
microcavity. In addition (iii), the theory enables us to consider the
non-equilibrium (hot) EP luminescence and opens a way for its observation in
organic-based nanodevices in analogy with the hot luminescence of molecules
and crystals \cite{Rebane_Saari78,Freiberg_Saari83}.

\section{Model Hamiltonian}

\label{sec:derivation} Consider an ensemble of molecules with two electronic
states ($1$ and $2$), which is described by the Hamiltonian
\begin{multline}
H_{0}=\hbar \sum_{m=1}^{\mathcal{N}}\Big[(\omega _{el}+\omega
_{0}X_{0m}^{2}+\sum\limits_{\sigma }\omega _{\sigma }X_{\sigma
m}^{2})b_{m}^{\dag }b_{m} 
-\omega _{0}X_{0m}(c_{0m}^{\dag }+c_{0m})b_{m}^{\dag }b_{m} \\
-\sum_{\sigma }\omega _{\sigma }X_{\sigma m}(c_{\sigma m}^{\dag }+c_{\sigma
m})b_{m}^{\dag }b_{m} 
+\omega _{0}c_{0m}^{\dag }c_{0m}+\sum\limits_{\sigma }\omega _{\sigma
}c_{\sigma m}^{\dag }c_{\sigma m}\Big],  \label{eq:H_02}
\end{multline}%
where $\omega _{el}$ is the frequency of the transition from the ground
state $1$ to the excited electronic state $2$, $\mathcal{N}$ is the number
of molecules. Annihilation of the excited state in the molecule $m$ is
described by the operator $b_{m}=\left\vert m1\right\rangle \langle m2|$,
and $b_{m}^{\dag }=\left\vert m2\right\rangle \langle m1|$ describes the
excitation in the $m$th molecule. The OA vibrations are accounted by the
annihilation and creation operators of the $i$th mode $c_{im}$ and $%
c_{im}^{\dag }$, respectively. A shift in the equilibrium position of the OA
vibration after excitation of the molecule $m$ is described by the quantity $%
X_{im}$ ($i=0,\sigma $). It is related to the standard Huang-Rhys factor $%
S_{im}$, e.g. $S_{0m}=X_{0m}^{2}$.

The light-matter interaction in the dipole approximation can be written
equivalently in two ways, by using $\mathbf{r\cdot E}$ or $\mathbf{p\cdot A}$
interaction Hamiltonians \cite{Scully97} with $\mathbf{E}$ being the electric field, $%
\mathbf{A}$ - the vector potential, and $\mathbf{p}$ - the canonical
momentum operator. Though both Hamiltonians for the light-matter interaction
are related via the gauge transformation and formally equivalent, they can
lead to different results \cite{Scully97}
and thus have to be used with a care. Since the exciton-polariton problem
was first formulated with the aid of the vector potential \cite%
{Hopfield58,Agranovich09} (see also \cite{Mukamel88OSA}), we also stick to
the $\mathbf{p\cdot A}$ Hamiltonian. In this case the electromagnetic-field
and the light-matter interaction Hamiltonians are {\small
\begin{equation}
H_{ph+int}=\hbar \sum_{\mathbf{q}}\omega _{\mathbf{q}}a_{\mathbf{q}}^{\dag
}a_{\mathbf{q}}-\frac{e}{Mc}\sum\limits_{m}\mathbf{A}(\mathbf{r}_{m})\cdot
\mathbf{p}(\mathbf{r}_{m}),  \label{eq:H_ph+int1}
\end{equation}%
} with $e$ and $M$ being the electron charge and mass, respectively, $c$ -
the light velocity, {\small
\begin{equation}
\mathbf{A}(\mathbf{r}_{m})=\sum_{\mathbf{q,e}_{\mathbf{q}}}\sqrt{\frac{2\pi
\hbar c^{2}}{V\omega _{\mathbf{q}}}}\mathbf{e}_{\mathbf{q}}\left[a_{\mathbf{q%
}}e^{i\mathbf{q\cdot r}_{m}}+ H.c.\right],  \label{eq:A}
\end{equation}%
}
\begin{equation}
\mathbf{p}(\mathbf{r}_{m})=-\frac{i\mathbf{D}M\omega _{el}}{e}%
(b_{m}-b_{m}^{\dag })  \label{eq:p}
\end{equation}%
Here $H.c.$ denotes Hermitian conjugate, $\mathbf{D}$ is the electronic
transition dipole moment, $\mathbf{e}_{\mathbf{q}}$ is the unit photon
polarization vector, $V$ is the photon quantization volume. In the
rotating-wave approximation, Hamiltonian $H_{ph+int}$ can be written as
\begin{equation}\label{eq:H_ph+int}
H_{ph+int}=\hbar \sum_{\mathbf{q}}\omega _{\mathbf{q}}a_{\mathbf{q}}^{\dag
}a_{\mathbf{q}} \\
-i\hbar \sum\limits_{m}\sum_{\mathbf{q}}\left[\frac{ge^{-i\mathbf{q\cdot r}
_{m}}}{\sqrt{\omega _{\mathbf{q}}}}b_{m}a_{\mathbf{q}}^{\dag }-H.c.\right],
\end{equation}
using Eqs.(\ref{eq:H_ph+int1}), (\ref{eq:A}) and (\ref{eq:p}). The
light-matter interaction parameter $g$ is
\begin{equation}
g=-\omega _{el}\sum_{\mathbf{e}_{\mathbf{q}}}\sqrt{\frac{2\pi }{\hbar V}}(%
\mathbf{e}_{\mathbf{q}}\cdot \mathbf{D})  \label{eq:g}
\end{equation}

Below we consider the LF vibrational subsystem as a classical one, since we
assume that $\hbar \omega _{\sigma }\ll k_{B}T$. An optical electronic
transition takes place at a fixed nuclear configuration, according to the
Franck-Condon principle. Then representing the electron-vibration coupling
quantity $\alpha _{m}=\sum_{\sigma }\omega _{\sigma }X_{\sigma m}(c_{\sigma
m}^{\dag }+c_{\sigma m})$ can be considered as a perturbation of the
electronic transition by nuclear motion. Essentially the LFOA vibrations \cite{Fai80} can
be thought of as a random modulation of the electronic transition frequency,
such that $\tilde{\omega}_{21}(t)=\omega _{el}+\omega _{st}/2-\alpha _{m}(t)$%
, where $\omega _{st}=2\sum_{\sigma }\omega _{\sigma }X_{\sigma m}^{2}=\hbar
K(0)/(k_{B}T)$ is the contribution of the LFOA vibrations to the Stokes
shift of the equilibrium absorption and luminescence spectra, and $\alpha
_{m}(t)$ is a random process. In that case $H_{0}$ becomes a stochastic
Hamiltonian and can be represented as
\begin{multline}
H_{0}(\alpha )=\hbar \sum\limits_{m}\Big\{\big[(\omega _{el}+\omega
_{st}/2+\omega _{0}X_{0m}^{2}-\alpha _{m}) \\
-\omega _{0}X_{0m}(c_{0m}^{\dag }+c_{0m})\big]b_{m}^{\dag }b_{m}+\omega
_{0}c_{0m}^{\dag }c_{0m}\Big\}  \label{eq:H_02HF(alpha)1}
\end{multline}%
We describe the electronic transition relaxation stimulated by LFOA
vibrations \cite{Muk95,Fai90OS,Fai93PR} by a one-parametric Gaussian process
with $\langle \alpha _{m}(t)\rangle _{1}=0$. In a particular case, this can
be a Gauss-Markov process with an exponential correlation function and a
characteristic attenuation time $\tau _{\sigma }$, $K_{m}(t)=\langle \alpha
_{m}(0)\alpha _{m}(t)\rangle _{1}=K(0)\exp (-|t|/\tau _{\sigma })$, where $%
\langle \cdot \rangle _{1}$ stands for the trace operation over the
reservoir variables in the ground electronic state $1$.

In this work we focus on calculation of the polariton luminescence spectrum,
bearing in mind two types of organic dye molecular systems: the H-aggregates
of dye molecules and solution of enhanced green fluorescent protein (eGFP).
The experimental \cite{Takazawa05} and the theoretical\cite{Fainberg19JPCC}
results show that the dipole-dipole interaction between the molecules
affects more strongly the absorption line-shape of the H-aggregates (due to
redistribution of intensities of vibronic transitions related to the HFOA
vibration) than their luminescence spectrum. Fast relaxation of the HFOA
vibration leads to the fact that only $0-0$ vibronic transition with respect
to the HFOA vibration contributes to the polariton luminescence (see below).
As for eGFP, the actual fluorophore of FPs is enclosed by a nano-cylinder
that consists of eleven $\beta $-sheets \cite{Gather_Yun14,Gather16}. This
protective shell acts as natural `bumper' and prevents close contact between
fluorophores of neighbouring FPs, limiting the intermolecular energy
migration even at the highest possible concentration. Thus, as a first
approximation, one can exclude the dipole-dipole interactions between
molecules from our consideration.

Hamiltonian $H_{0}(\alpha )$ can be diagonalized with respect to the HFOA
vibration using unitary Lang-Firsov transformation $\exp (-S)$ \cite%
{LangFirsov63,Silbey70}, where
\begin{equation}
S=\sum\limits_{m}X_{0m}(c_{0m}^{\dag }-c_{0m})b_{m}^{\dag }b_{m}
\label{eq:S}
\end{equation}%
As a result one get for the transformed Hamiltonian $\tilde{H}_{0}(\alpha
)=\exp (-S)H_{0}(\alpha )\exp (S)$ {\small
\begin{equation}
\tilde{H}_{0}(\alpha )=\hbar \sum\limits_{m}[(\omega _{el}+\omega
_{st}/2-\alpha _{m})b_{m}^{\dag }b_{m}+\omega _{0}c_{0m}^{\dag }c_{0m}]
\label{eq:H_02HF(alpha)2}
\end{equation}%
}

Next, the HFOA vibrations fast relaxation ($\sim 10-100$ $fs$) has to be
taken into account. We believe that the intramolecular relaxation related to
the HFOA vibrations takes place in a time shorter than the relaxation of the
LFOA system \cite{Fai99ACP,Fai02JCP}. Therefore, only the vibrationless
state $v=0$ with respect to the HFOA vibration will be populated in the
ground electronic state, and only the vibronic transitions $\left\vert
10\right\rangle \rightarrow \left\vert 2v\right\rangle $ can couple to light
to create polariton \cite{Rocca09}. In this case, the Hamiltonian $\tilde{H}%
_{0}(\alpha )$ can be rewritten as {\small
\begin{equation}
\tilde{H}_{0}(\alpha )=\hbar \sum\limits_{m}\sum\limits_{v}(\omega
_{2v}+\omega _{st}/2-\alpha _{m})b_{mv}^{\dag }b_{mv},
\label{eq:H_02HF(alpha)}
\end{equation}%
} where $\omega _{2v}=\omega _{el}+v\omega _{0}$, $b_{mv}^{\dag
}=|m2v\rangle \langle m10|$ and $b_{mv}=|m10\rangle \langle m2v|.$ For the
low excitations the operators $b_{mv}$ and $b_{mv}^{\dag }$ can be regarded
as bosons
\begin{equation}
\lbrack b_{m^{\prime }v^{\prime }},b_{mv}^{\dag }]=\delta _{mm^{\prime
}}\delta _{vv^{\prime }}  \label{eq:commutation_b_mov}
\end{equation}

In the new representation (Eq.~\ref{eq:H_02HF(alpha)}), the Hamiltonian of
the photons and of the light-matter interaction can be written in a compact
way by using the coherent states\cite{Glauber63},
\begin{equation}
\tilde{H}_{ph+int}=\hbar \sum_{\mathbf{q}}\omega _{\mathbf{q}}a_{\mathbf{q}%
}^{\dag }a_{\mathbf{q}} 
-i\hbar g\sum\limits_{m,\mathbf{q}}\left[ \frac{e^{-i\mathbf{q\cdot r} _{m}}%
}{\sqrt{\omega _{\mathbf{q}}}}\tilde{b}_{m}a_{\mathbf{q}}^{\dag }-\frac{ e^{i%
\mathbf{q\cdot r}_{m}}}{\sqrt{\omega _{\mathbf{q}}}}\tilde{b}_{m}^{\dag }a_{%
\mathbf{q}}\right],  \label{eq:H_ph+int_coh2}
\end{equation} 
where {\small
\begin{equation}
|2X_{0m}\rangle =e^{-X_{0m}^{2}/2}\sum\limits_{v}\frac{X_{0m}^{v}}{\sqrt{v!}}%
|m2v\rangle,  \label{eq:coher_state}
\end{equation}
\begin{equation}
\left(
\begin{array}{c}
\tilde{b}_{m} \\
\tilde{b}_{m}^{\dag }%
\end{array}%
\right) =e^{-X_{0m}^{2}/2}\sum\limits_{v=0}\frac{X_{0m}^{v}}{\sqrt{v!}}%
\left(
\begin{array}{c}
b_{mv} \\
b_{mv}^{\dag }%
\end{array}%
\right).  \label{eq:b_m}
\end{equation}%
}

\section{Polaritons Dispersion with Accounting of the Electron-Vibration
Interactions}

\label{sec:polaritons}

Usually, to calculate the polariton frequencies and the Hopfield
coefficients, one uses the electronic Hamiltonian \cite%
{Hopfield58,Knoester_Mukamel89}. In this case, however, the dispersion
equation for the polaritons cannot be reduced to the equation for the
transverse eigenmodes of the medium, $c^{2}q^{2}=\Omega _{s}^{2}\varepsilon
(\Omega _{s})$ \cite{Hau01,Fainberg18Advances,Fainberg19JPCC}, where $%
\varepsilon $ denotes the dielectric function. Below we perform averaging of
the Hopfield coefficients with respect to the LFOA vibrations ($\alpha _{m}$%
) directly. This enables us to get the dispersion equation that coincides
with the equation for the transverse eigenmodes of the medium. A side
benefit of taking the LFOA vibrations into account at calculations of the
polariton frequencies and the Hopfield coefficients we discuss in the
Supporting Information.

The bilinear Hamiltonian of the system
\begin{equation}
\tilde{H}=\tilde{H}_{0}(\alpha )+\tilde{H}_{ph+int}  \label{eq:H^tilde}
\end{equation}%
can be diagonalized by introducing the polariton operators $p_{s}$ as a
linear combination of operators $b_{mv}$ and $a_{\mathbf{q}}$ in the rotating
wave approximation
\begin{equation}
p_{s}=\sum_{\mathbf{q}}v_{s}(\mathbf{q)}a_{\mathbf{q}}+\sum%
\limits_{mv}u_{smv}b_{mv}.  \label{eq:p_sHF2}
\end{equation}%
Here $p_{s}$ denotes the annihilation operator for a polariton in branch $s$%
. The polariton operators have to obey the Bose commutation relations%
\begin{equation}
\lbrack p_{s},p_{s}^{\dag }]=\sum_{\mathbf{q}}|v_{s}(\mathbf{q)|}%
^{2}+\sum\limits_{mv}|u_{smv}|^{2}=1  \label{eq:commutation_psHF2}
\end{equation}%
The unknown coefficients $u_{smv}$ and $v_{s}(\mathbf{q)}$ has to be chosen
such, that the Hamiltonian (\ref{eq:H^tilde}) becomes diagonal in the
polariton operators,
\begin{equation}
\tilde{H}=\hbar \sum\limits_{s}\Omega _{s}p_{s}^{\dag }p_{s}
\label{eq:H(p)2}
\end{equation}%
The transformation coefficients $u_{smv}$ and $v_{s}(\mathbf{q)}$ obey the
following equations (see the Supporting Information)
\begin{equation}
u_{smv} =-i\zeta \lbrack \Omega _{s} -( \omega _{2v}+\omega _{st}/2)+\alpha
_{m}]v_{s}(\mathbf{q}) 
\times \sum_{\mathbf{q}}\frac{g}{\sqrt{\omega _{\mathbf{q}}}}e^{-X_{0}^{2}/2}%
\frac{X_{0}^{v}}{\sqrt{v!}}e^{-i\mathbf{q\cdot r}_{m}}  \label{eq:u_s(q)1}
\end{equation}
and%
\begin{equation}
(\Omega _{s}-\omega _{\mathbf{q}})v_{s}(\mathbf{q)}  
=\frac{ig}{\sqrt{\omega _{\mathbf{q}}}}\sum\limits_{mv}e^{-\frac{X_{0}^{2} }{%
2}+i\mathbf{q\cdot r}_{m}}\frac{X_{0}^{v}}{\sqrt{v!}}u_{smv},
\label{eq:v_s(q)4}
\end{equation}
where
\begin{equation}
\zeta [x]=\frac{P}{x } -i\pi \delta ( x),
\end{equation}
$P$ is the symbol of the principal value, and we omitted the subscript $m$
of the quantity $X_{0}$, assuming that all shifts are equal to each other, $%
X_{0m}\equiv X_{0}$.

Since $\alpha _{m}$ is a stochastic Gaussian variable, one can average the
amplitudes $u_{smv}(\mathbf{q})$ over the stochastic process using the
density matrix%
\begin{equation}
\rho _{11}^{(0)}\left( \alpha _{m}\right) =[2\pi K(0)]^{-1/2}e^{-\frac{%
\alpha _{m}^{2}}{2K(0)}}  \label{eq:rho_11equilibrium}
\end{equation}%
The averaging is reduced to the calculation of the integral
\begin{equation}
W_{av}(\Omega _{s}) =\frac{i}{\pi }\int_{-\infty }^{\infty }d\alpha \rho
_{11}^{(0)}\left( \alpha _{m}\right)  
\times \zeta \left[ \Omega _{s} -\left( \omega _{2v}+\frac{\omega _{st}}{2}%
\right)+\alpha _{m}\right]  \label{eq:Wav}
\end{equation}
representing the spectrum of $0\rightarrow v$ vibronic transition with
respect to the HFOA vibration \cite{Fainberg18Advances,Fainberg19JPCC}. The
imaginary part of "$-iW_{av}(\Omega _{s} ) $" with the minus sign, $-
\mathrm{Im}[-iW_{av}(\Omega _{s} ) ]=\mathrm{Re}W_{av}(\Omega _{s})$,
describes the absorption lineshape of the vibronic transition $0\rightarrow
v $, and the real part, Re$[-iW_{av}(\Omega _{s}\mathbf{)}]=\mathrm{Im}
W_{av}(\Omega _{s}\mathbf{)}$, describes the corresponding refraction
spectrum. For the \textquotedblright slow modulation\textquotedblright
limit, considered in this work, $K(0)\tau _{\sigma }^{2}>>1$, the quantity $%
W_{av}(\Omega _{s} ) $ is given by
\begin{equation}
W_{av}(\Omega _{s})=\sqrt{\frac{1}{2\pi K(0)}} \\
\times w\left(\frac{\Omega _{s}-(\omega _{2v}+\omega _{st}/2)}{\sqrt{2K(0)}}%
\right),  \label{eq:Wav_sm}
\end{equation}
where $w(z)=\exp (-z^{2})[1+i\mathrm{erfi}(z)]$ is the probability integral%
\cite{Abr64} of the complex argument $z$. This representation of the
function $W_{av}(\Omega _{s})$ is valid only in the vicinity of the central
frequency $\omega _{2v}+\omega _{st}/2$. The function $W_{av}(\Omega _{s} )$%
, however, can be calculated beyond the slow modulation limit \cite%
{Fainberg18Advances,Fainberg19JPCC} (see also \cite{Fai85}),
\begin{equation}
W_{av}(\Omega _{s}\mathbf{)}=\frac{\tau _{\sigma }}{\pi }\frac{\Phi
(1,1+x_{av};K(0)\tau _{\sigma }^{2})}{x_{av}}  \label{eq:Wav_hyper}
\end{equation}%
where $x_{av}=K(0)\tau _{\sigma }^{2}+i\tau _{\sigma }(\omega _{2v}+\omega
_{st}/2-\Omega _{s})$, and $\Phi (1,1+x_{av};K(0)\tau _{\sigma }^{2})$ is
the confluent hypergeometric function \cite{Abr64}. It is worthy to note
that Eq.(\ref{eq:Wav_hyper}) is valid for both small and large detuning away
from the central frequency. Note, that Eq.(\ref{eq:Wav_hyper}) is used in
our numerical calculations.

The averaged value of $u_{smv}$ can be obtained using Eqs.(\ref{eq:u_s(q)1}%
), (\ref{eq:v_s(q)4}) and (\ref{eq:Wav}), it is
\begin{equation}
\bar{u}_{smv}(\mathbf{q})=\int u_{smv}(\mathbf{q})\rho _{11}^{(0)}\left(
\alpha _{m}\right) d\alpha _{m}  
=-\frac{\pi g }{\sqrt{\omega _{\mathbf{q}}}} e^{-X_{0}^{2}/2-i\mathbf{q\cdot
r}_{m}}\frac{X_{0}^{v}}{\sqrt{v!}}W_{av}[\Omega _{s}(\mathbf{q)}]v_{s}(%
\mathbf{q)}  \label{eq:u_smv(q)average}
\end{equation}

Finally, we can find the relation for the polariton frequencies $\Omega _{s}$
and the amplitudes $v_{s}(\mathbf{q})$ by substitution of $\bar{u}_{smv}$
from Eq.(\ref{eq:u_smv(q)average}) into Eq.(\ref{eq:v_s(q)4}),
\begin{equation}
(\Omega _{s}-\omega _{\mathbf{q}})v_{s}(\mathbf{q)}=-\frac{ig^{2}}{\sqrt{%
\omega _{\mathbf{q}}}}\pi W_{a}(\Omega _{s})  
\times\sum_{\mathbf{q}^{\prime }}\frac{v_{s}(\mathbf{q}^{\prime })} {\sqrt{%
\omega_{\mathbf{q}}^\prime}}\sum\limits_{m}\exp [i(\mathbf{q}-\mathbf{q}%
^{\prime })\mathbf{\cdot r}_{m}],  \label{eq:polariton_dispersion5}
\end{equation}
where%
\begin{equation}
W_{a}(\Omega _{s})=e^{-X_{0}^{2}}\sum\limits_{v}\frac{X_{0}^{2v}}{v!}%
W_{av}(\Omega _{s})  \label{eq:Wa2}
\end{equation}%
is the equilibrium molecular spectrum in the presence of both HF and LF OA
vibrations.

For the systems without the spatial symmetry (like eGFP molecules in the
diluted protein solution\cite{Gather_Yun14}) the photon amplitude $v_{s}(%
\mathbf{q})$ enters the right-hand side of Eq.(\ref{eq:polariton_dispersion5}%
) with all possible values of its argument $\mathbf{q}^{\prime }$. The
values of $\mathbf{q}^{\prime }$, which differ from the fixed value $\mathbf{%
q}$ in Eq.(\ref{eq:polariton_dispersion5}), describe the light scattering
from the inhomogeneities of the medium. When this scattering is neglected%
\cite{Agranovich03}, only the term with $\mathbf{q}^{\prime }=\mathbf{q}$
survives under the sum sign, and the sum on the right-hand side of Eq.(\ref%
{eq:polariton_dispersion5}) equals to $\sum_{m}\exp [i(\mathbf{q-q^{\prime
})\cdot r}_{m}]=\mathcal{N\delta }_{\mathbf{qq}^{\prime }}$. In this
zero-order approximation, Eq.(\ref{eq:polariton_dispersion5}) turns into the
dispersion equation for the polaritons
\begin{equation}
\Omega _{s}(\mathbf{q)-}\omega _{\mathbf{q}}=-\frac{ig^{2}\mathcal{N}}{%
\omega _{\mathbf{q}}}\pi W_{a}[\Omega _{s}(\mathbf{q)}]
\label{eq:polariton_dispersionHF}
\end{equation}%
From the Eq.(\ref{eq:g}) we have the following relation for the parameters
\begin{equation}
g^{2}\mathcal{N}=\omega _{el}^{2}Q,  \label{eq:Q1}
\end{equation}%
where $Q=4\pi N\frac{\mathbf{D}_{12}\mathbf{D}_{21}}{\hbar }$ (note that in
Refs.\cite{Fainberg18Advances,Fainberg19JPCC} $Q$ corresponds to the
parameter $q$), and $N=\mathcal{N}/V$ is the density of molecules. At the
derivation of Eq.(\ref{eq:Q1}), we summed up over two independent electric
field polarization directions\cite{Scully97} $\mathbf{e}_{\mathbf{q}}$ for
each $\mathbf{q}$. This summation cancels the factor $1/2$, which
otherwise would appear in Eq.(\ref{eq:Q1}).

In the case without the symmetry and when the scattering from the
inhomogeneities is neglected, the polariton frequency $\Omega _{s}$ and the
exciton amplitude $u_{smv}$ become functions of $\mathbf{q}$, $\Omega
_{s}=\Omega _{s}(\mathbf{q})$ and $u_{smv}=u_{smv}(\mathbf{q})$, and one can
introduce the polariton operators $p_{s\mathbf{q}}$
\begin{equation}
p_{s\mathbf{q}}=v_{s}(\mathbf{q)}a_{\mathbf{q}}+\sum\limits_{mv}u_{smv}(%
\mathbf{q)}b_{mv}  \label{eq:p_sqHF}
\end{equation}%
where $p_{s\mathbf{q}}$ denotes the annihilation operator of a polariton
with wave vector $\mathbf{q}$ corresponding to the branch $s$. Then Eq.(\ref%
{eq:H(p)2}) takes the form
\begin{equation}
\tilde{H}=\hbar \sum\limits_{s\mathbf{q}}\Omega _{s}(\mathbf{q)}p_{s\mathbf{q%
}}^{\dag }p_{s\mathbf{q}},  \label{eq:H(p)3}
\end{equation}%
where the polariton operators $p_{s\mathbf{q}}$ obey the Bose commutation
relation%
\begin{equation}
\lbrack p_{s\mathbf{q}},p_{s\mathbf{q}}^{\dag }]=|v_{s}(\mathbf{q)|}%
^{2}+\sum\limits_{mv}|u_{smv}(\mathbf{q)}|^{2}=1
\label{eq:commutation_psqHF}
\end{equation}

The averaged dispersion equation for the polaritons, Eq.(\ref%
{eq:polariton_dispersionHF}), can be reduced to the equation for the
transverse eigenmodes of the medium, $c^{2}q^{2}=\Omega _{s}^{2}\varepsilon
(\Omega _{s})$, \cite{Hau01,Fainberg18Advances,Fainberg19JPCC}. Indeed,
bearing in mind that $\omega _{\mathbf{q}}=cq/\sqrt{\varepsilon _{0}}$,
where $n_{0}=\sqrt{\varepsilon _{0}}$ is the background refraction index of
the medium, in the rotating-wave approximation when $\Omega _{s}\omega _{%
\mathbf{q}}\approx \Omega _{s}^{2}$ and $\omega _{el}^{2}/\Omega
_{s}^{2}\approx 1$ one gets from Eqs.(\ref{eq:polariton_dispersionHF}) and (%
\ref{eq:Q1})
\begin{equation}
c^{2}q^{2}/\Omega _{s}^{2}=\varepsilon _{0}[1+Qi\pi W_{a}(\Omega _{s})]
\label{eq:Dispersion4}
\end{equation}%
The right-hand side of Eq.(\ref{eq:Dispersion4}) presents the dielectric
function $\varepsilon (\Omega _{s})$ in the absence of the dipole-dipole
interaction between the molecules \cite{Fainberg18Advances,Fainberg19JPCC}.
It is worthy to note that the dispersion equation for the polaritons can be
recast in the form $c^{2}q^{2}=\Omega _{s}^{2}\varepsilon (\Omega _{s})$
also when the non-resonant terms and the $\mathbf{A}^{2}$ term in the
light-matter interaction part of the Hamiltonian $H_{ph+int}$ are also taken
into account \cite{Hopfield58,Mukamel88OSA,Knoester_Mukamel91}. Therefore,
since both Eq.(\ref{eq:polariton_dispersionHF}) and Eq.(\ref{eq:Dispersion4}%
) have the same accuracy in the rotating-wave approximation, below in our
calculations we stick to Eq.(\ref{eq:Dispersion4}). Eq.(\ref{eq:Dispersion4}%
) can be obtained from Eq.(\ref{eq:polariton_dispersionHF}) under the
following substitutions
\begin{equation}
\omega _{\mathbf{q}}\rightarrow \omega _{\mathbf{q}}+\Omega _{s},\quad
\omega _{el}^{2}\rightarrow \Omega _{s}^{2}.  \label{eq:substitution1}
\end{equation}%
It is worthy to note that this substitution, Eq.(\ref{eq:substitution1}), is
consistent with the accounting of the non-resonant terms.

To take into account the non-resonant contributions to the HFOA $%
0\rightarrow v$ vibronic transition spectrum $W_{av}(\Omega _{s})$, Eqs.(\ref%
{eq:Wav_sm}) and (\ref{eq:Wav_hyper}), we use the following arguments\cite%
{Briggs06PRL,Fainberg18Advances}. Taking the spectrum $W_{av}(\Omega _{s})$
to be centred at the frequency $\Omega _{s}=\omega _{2v}+\omega _{st}/2$ and
having a finite width $\Gamma$, we derive from the dispersion relation Eq.(%
\ref{eq:Dispersion4}) the resonant contribution to the spectrum $%
W_{av}(\Omega _{s})\sim (i/\pi )/[\Omega _{s}-(\omega _{2v}+\omega _{st}/2)]$
for the case $|\Omega _{s}-(\omega _{2v}+\omega _{st}/2)|>>\Gamma $. Thus to
extend the validity of the spectral function $W_{av}(\Omega_{s})$ in the
non-resonant approximation, it is enough to add the non-resonant term $%
(-i/\pi )/[\Omega _{s}+(\omega _{2v}+\omega _{st}/2)]$ to it, i.e. {\small
\begin{equation}
W_{av}^{non}(\Omega _{s}\mathbf{)}=W_{av}(\Omega _{s})-\frac{i/\pi }{\Omega
_{s}+(\omega _{2v}+\omega _{st}/2)}  \label{eq:Wav_non}
\end{equation}%
} Accordingly, the equilibrium molecular spectrum $W_{av}(\Omega _{s})$, Eq.(%
\ref{eq:Wa2}), on the right-hand side of the dispersion equation, Eq.(\ref%
{eq:Dispersion4}), should be replaced by $W_{a}^{non}(\Omega _{s})${\small
\begin{equation}
W_{a}^{non}(\Omega _{s})=\exp (-X_{0}^{2})\sum\limits_{v}\frac{X_{0}^{2v}}{v!%
}W_{av}^{non}(\Omega _{s})  \label{eq:Wa2non}
\end{equation}%
}

\begin{figure}
\centering
\includegraphics[scale=0.40]{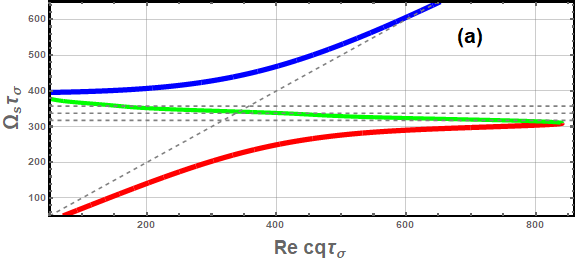} %
\includegraphics[scale=0.40]{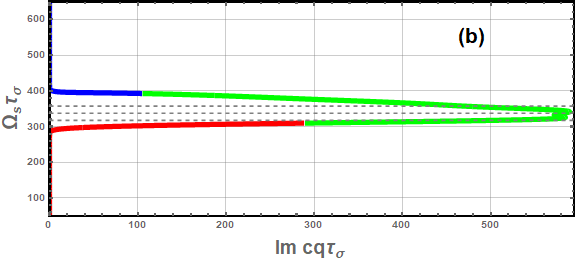}
\caption{{\small The polariton dispersion curves, $\Omega _{s}(\mathbf{q})$, (eq.~\ref{eq:Dispersion4}) for real (a) and imaginary (b) parts of the wavevector $q$. The red/blue curves correspond to the lower/upper polariton branches, the green curve is the leaky part. The model parameters are obtained by fitting the TC dye
experimental absorption spectrum in Ref.\cite{Fainberg19JPCC}: $K(0)%
\tau _{\sigma }^{2}=80$, $Q\tau _{\sigma %
}=84 $, $\omega _{st}\tau _{\sigma }=28.6$, $%
\omega _{0}\tau _{\sigma }=20$, $X_{0}^{2}=0.454$, $%
\omega _{el}\tau _{\sigma }=303.2$, $1/\tau _{%
\sigma }=75\,cm^{-1}$.}}
\label{fig:Dispersion1}
\end{figure}

In Fig.\ref{fig:Dispersion1} we plot the EP dispersion curves, Eqs.(\ref%
{eq:Dispersion4}) and (\ref{eq:Wav_non}), with the model parameters tuned to the
experimental observation of the polariton spectrum of the TC dye molecules
nanofiber \cite{Takazawa10}. To satisfy Eqs.(\ref{eq:Dispersion4}), the
wavevector $q$ has to have a non-zero imaginary part, $q=q^{\prime}+iq^{%
\prime \prime}$. The imaginary part of wavevector $q^{\prime\prime }$ as a
function of the polariton frequency is plotted in Fig.\ref{fig:Dispersion1}
b. In the leaky part of the obtained polariton dispersion, i.e. in the range
of frequencies corresponding to the gap between the lower and the upper
polariton branches (green line), the value of $q^{\prime\prime}$ reaches its
largest values. It can be considered as a polariton absorption spectrum.

Note that Eq.(\ref{eq:Dispersion4}) can be further extended to the case of
the intermolecular dipole-dipole interaction with the help of the mean-field
approach\cite{Fainberg18Advances,Fainberg19JPCC}. The intermolecular
dipole-dipole interaction, however, does not change the general behaviour of
the dispersion curves. This supports the approach used in our paper.

\begin{figure}
\centering
\includegraphics[scale=0.40]{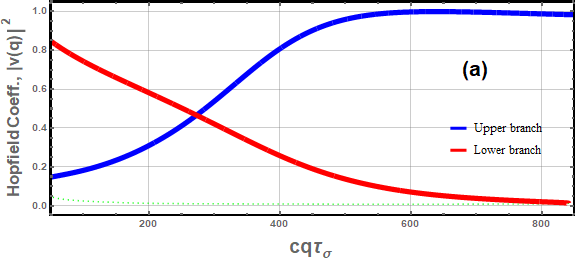} %
\includegraphics[scale=0.40]{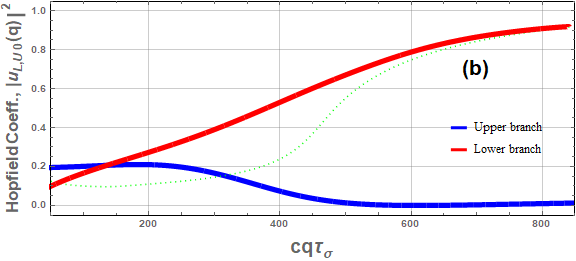}
\caption{ {\small The Hopfield coefficients as functions of the
wavevector $q=\left\vert \bm q\right\vert$. (a) $|v_{L}(\bm q)|^{2}$ (red
line) and $|v_{U}(\bm q)|^{2}$ (blue line); (b) coefficients $%
\sum_{m}|u_{Lm0}(\bm q)|^{2}$ (red line) and $\sum_{m}|u_{Um0}(\bm %
q)|^{2}$. Other parameters are identical to those as in Fig.~\ref%
{fig:Dispersion1}.}}
\label{fig:Hopfield1}
\end{figure}

With the help of Eqs.(\ref{eq:commutation_psqHF}), (\ref{eq:u_smv(q)average}%
), (\ref{eq:Dispersion4}) and (\ref{eq:substitution1}) we can derive the
Hopfield coefficients {\small
\begin{equation}
|v_{s}(\mathbf{q})|^{2}=\Bigg[1 +\frac{\pi ^{2}\Omega _{s}^{2}(\mathbf{q)}Q}{%
\omega _{\mathbf{q}}+\Omega _{s}(\mathbf{q})}\exp (-X_{0}^{2})  
\times\sum_{v}\frac{X_{0}^{2v}}{v!}\left\vert W_{av}^{non}[\Omega _{s}(%
\mathbf{q})]\right\vert ^{2}\Bigg] ^{-1}  \label{eq:v_s(q)^2}
\end{equation}
\begin{equation}
\sum_{m}|\bar{u}_{smv}(\mathbf{q})|^{2}=|v_{s}(\mathbf{q})|^{2}\frac{\pi
^{2}\Omega _{s}^{2}(\mathbf{q})Q}{\omega _{\mathbf{q}}+\Omega _{s}(\mathbf{q})%
} \\
\times\exp (-X_{0}^{2})\frac{X_{0}^{2v}}{v!}|W_{av}^{non}[\Omega _{s}(%
\mathbf{q})]|^{2}  \label{eq:u_smv(q)average^2}
\end{equation}
} The Hopfield coefficients calculated by Eqs.(\ref{eq:v_s(q)^2}) and (\ref%
{eq:u_smv(q)average^2}) are plotted for the upper (U) and the lower (L)
polariton branches as a functions of the wavevector $\bm q$ absolute value
in Fig.\ref{fig:Hopfield1}.

The obtained Hopfeld coefficients, Eqs.(\ref{eq:v_s(q)^2}) and (\ref%
{eq:u_smv(q)average^2}), satisfy the standard normalizing conditions.

One can see from Eq.(\ref{eq:u_smv(q)average}) that%
\begin{equation}
\bar{u}_{smv}(\mathbf{q})=\frac{\bar{u}_{sv}(\mathbf{q)}}{\sqrt{\mathcal{N}}}%
\exp (-i\mathbf{q\cdot r}_{m})  \label{eq:u_smv(q)average2}
\end{equation}%
where%
\begin{equation}
\bar{u}_{sv}(\mathbf{q})=- \frac{\pi g\sqrt{\mathcal{N}}}{\sqrt{\omega _{%
\mathbf{q}}}}e^{-X_{0}^{2}/2}\frac{X_{0}^{v}}{\sqrt{v!}}W_{av}[\Omega _{s}(%
\mathbf{q})]v_{s}(\mathbf{q})  \label{eq:u_sv(q)2}
\end{equation}%
does not depend on $m$. Using Eq.(\ref{eq:commutation_psqHF}), we derive
\begin{equation}
\lbrack p_{s\mathbf{q}},p_{s\mathbf{q}}^{\dag }]=|v_{s}(\mathbf{q)|}%
^{2}+\sum\limits_{v}|\bar{u}_{sv}(\mathbf{q)}|^{2}=1
\label{eq:commutation_psqHFsym}
\end{equation}%
with
\begin{equation}
|\bar{u}_{sv}(\mathbf{q)|}^{2}=\sum_{m}|\bar{u}_{smv}(\mathbf{q})|^{2}.
\label{eq:u_smv(q)average^2-u_sv(q)average^2}
\end{equation}

\section{Polariton Luminescence Spectrum}

\subsection{General formulas}

\begin{figure}[tbp]
\centering
\includegraphics[scale=0.38]{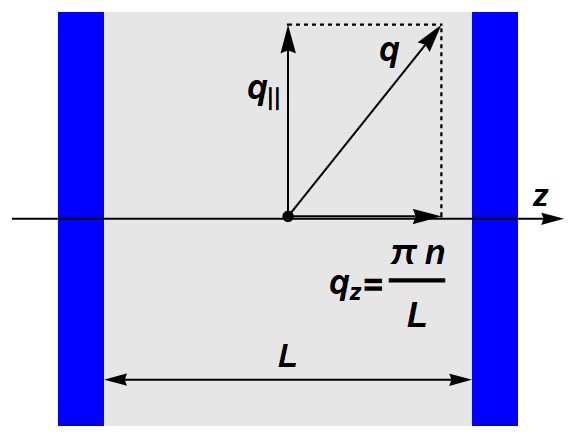}
\caption{{\small The molecular substance is located between the
cavity mirrors separated by distance \textit{L}. The in-plane wave vector $
\mathbf{q}_{||}$ and its $Z$-component are shown in the figure.}}
\label{fig:cavity}
\end{figure}

The frequency spectrum of a light emitting system, {\small
\begin{equation}
I(\omega )=2\mathrm{Re}\int_{0}^{\infty }d\tau \langle \mathbf{E}^{(-)}(t)%
\mathbf{E}^{(+)}(t-\tau )\rangle e^{-i\omega \tau }  \label{eq:I(omega)}
\end{equation}%
} is generally calculated from the two-time correlation function of the
quantized electric field \cite{Eberly77}, $\mathbf{\hat{E}}(\mathbf{r},t)=%
\mathbf{E}^{(+)}(t)+\mathbf{E}^{(-)}(t)$, obtained by the photon leakage
(see coefficient $\eta _{phot}$ below) through the mirrows of a microcavity
(see Fig.\ref{fig:cavity}).

The quantum averaging $\langle \dots\rangle $ indicates taking the trace
over the density matrices of the phonon and polariton subsystems. Since we consider the LF vibrational system as a classical one, this procedure includes also stochastic averaging (see below).
To relate this out field with intracavity one, we will use the quasimode approximation~\cite{Savona99}, according to which the in-out coupling conserves the in-plane wave vector $\mathbf{q}_{||}$. Then one can use $\mathbf{q}_{||}$
to denote external (emitted) photons \cite{Zoubi_Rocca05,Lidzey08,Rocca09},
and
\begin{equation}
\mathbf{E}^{(+)}=[\mathbf{E}^{(-)}]^{\dag }=i\sqrt{2\pi \hbar \omega _{%
\mathbf{q}_{||}}}\bar{a}_{\mathbf{q}_{||}}u_{\mathbf{q}_{||}}(\mathbf{r)}%
e^{-i\omega _{\mathbf{q}_{||}}t}  \label{eq:E_quantum1}
\end{equation}%
where $\bar{a}_{\mathbf{q}_{||}}$ is the external photon annihilation
operator, the amplitude $u_{\mathbf{q}_{||}}(\mathbf{r})$ describes the
field spatial dependence. Substituting Eq.(\ref{eq:E_quantum1}) into Eq.(\ref%
{eq:I(omega)}), we obtain
\begin{equation}
I(\omega )=4\pi \hbar \omega _{\mathbf{q}_{||}}|u_{\mathbf{q}_{||}}(\mathbf{%
r)|}^{2} 
\times \mathrm{Re}\int_{0}^{\infty }d\tau \left\langle \bar{a}_{\mathbf{q}%
_{||}}^{\dag }(t)\bar{a}_{\mathbf{q}_{||}}(t-\tau )\right\rangle e^{-i\omega
\tau }  \label{eq:I(omega)2}
\end{equation}

Consider the polariton luminescence in a single-mode cavity with the
eigenfrequency
\begin{equation}
\omega _{\mathbf{q}n}=cq/n_{0}=\frac{c}{n_{0}}\sqrt{q_{||}^{2}+q_{z}^{2}}
\label{eq:frequency}
\end{equation}
where

\begin{equation}
\mathbf{q=(}q\mathbf{_{||},}q_{z}\mathbf{)}  \label{eq:q}
\end{equation}
is the total wave vector, $n_{0}$ is the background refraction index of the
medium between the mirrors, $q_{z}=n\pi /L$ ($n=1,2,3,...$) and $q=|\mathbf{q%
}|$. The intracavity photon and the exciton operators can be expressed in
terms of the polariton operators by means of the inverse polariton
transformation \cite{Tyablikov67,Knoester_Mukamel91}
\begin{equation}
a_{\mathbf{q}}=\sum\limits_{s}v_{s}^{\ast }(\mathbf{q)}p_{s\mathbf{q}%
},b_{mv}=\sum\limits_{s\mathbf{q}}u_{smv}^{\ast }(\mathbf{q})p_{s\mathbf{q}},
\label{eq:a_q,b_mv}
\end{equation}%
From Eq.(\ref{eq:I(omega)2}) using Eq.(\ref{eq:a_q,b_mv}) we get
\begin{equation}
I_{p\mathbf{q}_{||}}(\omega )=4\pi \eta _{phot}\sum_{ss^{\prime }}\hbar
\omega _{\mathbf{q}_{||}}|u_{\mathbf{q}_{||}}(\mathbf{r)|}^{2}\mathrm{Re}
\int_{0}^{\infty }d\tau 
 e^{-i\omega \tau }\left\langle v_{s}(\mathbf{q)}v_{s^{\prime }}^{\ast
}(\mathbf{q)}p_{s\mathbf{q}}^{\dag }(t)p_{s^{\prime }\mathbf{q}}(t-\tau
)\right\rangle,  \label{eq:I(omega)3}
\end{equation}
where both the amplitudes $v_{s}(\mathbf{q})$ and the polariton variables
has to be averaged over the phonon bath.

In the article we consider strong elecron-vibrational interaction. Therefore
this interaction has been taken into account in the previous section at the
calculation of the amplitudes $v_s(\mathbf{q})$ and $u_{smv}(\mathbf{q})$ .
Moreover the amplitudes were also averaged over the variables $\alpha_m$,
which are describing the influence of the LFOA vibrations. This enabled us
to reduce the obtained polariton dispersion equation to the dispersion
relation known from the dielectric theory of polaritons\cite{Hau01}. The
latter indicates correctness of the averaging procedure. Therefore, since
the averaging of the amplitudes $v_s(\mathbf{q})$ and $u_{smv}(\mathbf{q})$
has been done, we can factorize the expectation value
\begin{equation}
\left\langle v_{s}(\mathbf{q)}v_{s^{\prime }}^{\ast }(\mathbf{q)}p_{s\mathbf{%
q}}^{\dag }(t)p_{s^{\prime }\mathbf{q}}(t-\tau )\right\rangle 
=\langle v_{s}(\mathbf{q)}v_{s^{\prime }}^{\ast }(\mathbf{q)\rangle \langle }%
p_{s\mathbf{q}}^{\dag }(t)p_{s^{\prime }\mathbf{q}}(t-\tau )\rangle ,
\label{eq:factorization}
\end{equation}
Additional arguments in favour of such factorization will be given in the next section.

\subsection{Calculation of the polariton two-particle expectation value}

\label{sec:two-particle} In this section we calculate the polariton
two-particle expectation value, Eq.(\ref{eq:factorization}). The polariton
operators were defined above using the thermal averaging of the LFOA
vibrations with respect to the ground electronic state. Usually the
polariton excitation occurs in the range of energies where the polaritons
are excitons, and therefore the excitation is followed by relaxation of the
vibrational subsystem in the excited electronic state. For this reason,
under the averaging $\langle \mathbf{...}\rangle $ in the term $\langle p_{s%
\mathbf{q}}^{\dag }(t)p_{s^{\prime }\mathbf{q}}(t-\tau )\rangle $ on the
right-hand side of Eq.(\ref{eq:factorization}) we understand the trace over
the phonon bath density matrix in the excited electronic state. It is worthy
to note that the factorization, Eq.(\ref{eq:factorization}), can be
justified as follows. The amplitudes $v_{s}(\mathbf{q)}$ and $\bar{u}_{smv}(%
\mathbf{q)}$ (or $\bar{u}_{sv}(\mathbf{q)}$) are calculated using the
averaging with respect to the ground electronic state (see the section
above). At the same time, $\langle p_{s\mathbf{q}}^{\dag }(t)p_{s^{\prime }%
\mathbf{q}}(t-\tau )\rangle $ is calculated for the emission involving
relaxation in the excited electronic state. Between these events a rapid
dephasing occurs, in particular, due to the HFOA vibrations. Therefore, the
averages of the amplitudes $v_{s}(\mathbf{q)}$, $u_{smv}(\mathbf{q)}$ ($%
u_{sv}(\mathbf{q)}$) and $\langle p_{s\mathbf{q}}^{\dag }(t)p_{s^{\prime }%
\mathbf{q}}(t-\tau )\rangle $ can be carried out separately.

We rewrite the $\alpha _{m}$-dependent part of the Hamiltonian $\tilde{H}%
_{0}(\alpha )$, Eq.(\ref{eq:H_02HF(alpha)}), the term $\tilde{H}_{0}^{\prime
}(\alpha )\equiv -\hbar \sum_{m}\sum_{v=0}\alpha _{m}b_{mv}^{\dag }b_{mv}$,
through the polariton variables. Using Eqs.(\ref{eq:p_sqHF}), (\ref%
{eq:u_sv(q)2}) and (\ref{eq:a_q,b_mv}), we write
\begin{equation}
\tilde{H}_{0}^{\prime }(\alpha )=-\hbar
\sum\limits_{m}\sum\limits_{v=0}\alpha _{m}b_{mv}^{\dag }b_{mv} 
=-\hbar \sum_{\mathbf{qq}^{\prime }}\alpha (\mathbf{q}-\mathbf{q}^{\prime
})\sum\limits_{v=0,s,s^{\prime }}\bar{u}_{sv}(\mathbf{q})\bar{u}_{s^{\prime
}v}^{\ast }(\mathbf{q}^{\prime })p_{s\mathbf{q} }^{\dag }p_{s^{\prime }
\mathbf{q}^{\prime }},  \label{eq:alpha_mb^+_mvb_mv3}
\end{equation}
where
\begin{equation}
\alpha (\mathbf{q}-\mathbf{q}^{\prime } )= \frac{1}{\mathcal{N}}%
\sum\limits_{m}\alpha _{m}\exp [-i(\mathbf{q}-\mathbf{q}^{\prime })\cdot
\mathbf{\ r}_{m}].  \label{eq:alpha(q-q')}
\end{equation}

The polariton operators obey the Heisenberg equations $\frac{\partial p_{s%
\mathbf{q}}(t)}{\partial t} =\frac{i}{\hbar }[\tilde{H}+\tilde{H}%
_{0}^{\prime }(\alpha ),p_{s\mathbf{q}}(t)]$, which explicit form is
\begin{equation}\label{eq:dp_sk/dt}
\frac{\partial p_{s\mathbf{q}}(t)}{\partial t} =-i\Omega _{s}(\mathbf{q})p_{s%
\mathbf{q}}(t)+  
+i\sum\limits_{v\mathbf{q}^{\prime }s^{\prime }}\bar{u}_{sv}(\mathbf{q})\bar{%
u}_{s^{\prime }v}^{\ast }(\mathbf{q}^{\prime })\alpha (\mathbf{q}-\mathbf{q}%
^{\prime } ) p_{s^{\prime }\mathbf{q}^{\prime }}(t).
\end{equation}
Here we have used Bose commutation relations for the polaritons.

Since other OAHF vibration states with $v\geqslant 1$ relax very fast (see
discussion above), only the vibrationless state $v=0$ with respect to the
HFOA vibration will contribute to the right-hand side of Eq.(\ref%
{eq:dp_sk/dt}). The last term on the right-hand side of Eq.(\ref{eq:dp_sk/dt}%
) describes coupling between different $\mathbf{q}$ and $s$ components of
the polariton operator, resulting from the vibrational perturbation $\alpha (%
\mathbf{q}-\mathbf{q}^{\prime })$. This leads to both transitions between
different polariton branches and moving of the corresponding wave packet
along the dispersion curve. Note that presence of the HFOA vibrations
generates a number of different polariton branches related to either $U$ or $%
L$ family (see the Supporting Information). However, for a weak
electron-vibrational coupling with respect to the HFOA vibration, i.e. $%
X_{0m}<1$, increase of the vibronic index $v$ increases the energy of the
corresponding polariton branches. In other words, the energy increase with
the increase of $v$ is conserved also for the polariton branches with $%
X_{0m}<1$. The case $X_{0m}<1$ is realized in the majority of organic dyes.

\subsection{Single mode cavity and semiclassical approximation}

\label{sec:single-mode}

Usually they measure a luminescence signal at a specific angle corresponding
to the specific values of $\mathbf{q}_{||}$ and $\mathbf{q}=( q _{||}, q_{z})
$, respectively. So, in the single-mode cavity we are interested in a
specific value of $\mathbf{q}$. Assuming that $\mathbf{q}=\mathbf{q}^{\prime
}$ in Eq.(\ref{eq:dp_sk/dt}), we get
\begin{equation}
\frac{\partial \tilde{p}_{s\mathbf{q}}(t)}{\partial t}=i\alpha (0\mathbf{)}%
\sum\limits_{s^{\prime }\neq s}\bar{u}_{s0}(\mathbf{q})\bar{u}_{s^{\prime
}0}^{\ast }(\mathbf{q})\tilde{p}_{s^{\prime }\mathbf{q}}e^{i(d_{s\mathbf{q}%
}-d_{s^{\prime }\mathbf{q}})t}  \label{eq:dp^tilda_sk/dt}
\end{equation}%
where we put $p_{s\mathbf{q}}(t)=\tilde{p}_{s\mathbf{q}}\exp (-id_{s\mathbf{q%
}}t)$ and denoted%
\begin{equation}
d_{s\mathbf{q}}=\Omega _{s}(\mathbf{q)-}\alpha (0)|\bar{u}_{s0}(\mathbf{q}%
)|^{2}.  \label{eq:d_sk}
\end{equation}%
Eq.(\ref{eq:dp^tilda_sk/dt}) includes the last term on the right-hand side
of Eq.(\ref{eq:dp_sk/dt}) only at the largest value of the vibrational
perturbation $\alpha (\mathbf{q}-\mathbf{q}^{\prime })$ for $\mathbf{q}%
^{\prime }=\mathbf{q}$, $\alpha (0\mathbf{)}$. The corresponding processes
for $\mathbf{q}^{\prime }\neq \mathbf{q}$ describe polariton relaxation
along the dispersion curve and result in the change of the population of
polaritons with wave vector $\mathbf{q}$. The latter rather effects the
luminescence intensity $I_{p\mathbf{q}_{||}}(\omega )$ than change its
spectrum in\ the slow modulation limit under consideration in this section,
and will be studied elsewhere.

To solve the problem, first, we use the semiclassical approximation \cite%
{Per74} considering the quantity $\alpha _{m}$ as time-independent.
Physically, the approximation is applicable for the slow modulation limit
and means that only vertical (Frank-Condon) electron-vibrational transitions take place.
Generalization to the case of arbitrary modulation and accounting for the
non-vertical transitions will be carried out below in assumption of the
Gauss-Markov process.

In the semiclassical approximation we have a couple of Eqs.(\ref%
{eq:dp^tilda_sk/dt}) written for each polariton branch $s,s^{\prime }=U,L$.
Solving them for a specific case of equal initial excitation of both
polariton branches, with the help of Eq.(\ref{eq:I(omega)3}) in the
steady-state regime ($t=0$) we get
\begin{multline}
I_{p\mathbf{q}_{||}}(\omega )=4\pi \eta _{phot}\hbar \omega _{\mathbf{q}%
_{||}}|u_{\mathbf{q}_{||}}(\mathbf{r)|}^{2}\mathrm{Re}\int_{0}^{\infty
}d\tau e^{-i\omega \tau } \\
\times \Bigg\langle e^{i\frac{d_{U\mathbf{q}}+d_{L\mathbf{q}}}{2}\tau
}\sum_{s=U,L}\Big[\langle |v_{s}(\mathbf{q)|}^{2}\mathbf{\rangle }P_{ss}(%
\mathbf{q,}\tau )  
+\langle v_{s}(\mathbf{q)}v_{s^{\prime }}^{\ast }(\mathbf{q)\rangle }%
P_{ss^{\prime }}(\mathbf{q,}\tau )\Big]\Bigg\rangle
\label{eq:I(omega)Fourier}
\end{multline}%
The diagonal element $P_{ss}(\mathbf{q,}\tau )$ is expressed by the formula
\begin{equation}
P_{ss}(\mathbf{q,}\tau )=\frac{|\tilde{p}_{\mathbf{q}}(0)|^{2}}{2}   \sum_{+,-}\left[ 1\pm \frac{2\alpha (0\mathbf{)}\bar{u}_{U0}\bar{u}%
_{L0}+d_{s\mathbf{q}}-d_{s^{\prime }\mathbf{q}}}{\sqrt{(d_{L\mathbf{q}}-d_{U%
\mathbf{q}})^{2}+4\bar{c}}}\right] 
\times e^{\mp \frac{i\tau }{2}\sqrt{(d_{L\mathbf{q}}-d_{U\mathbf{q}})^{2}+4%
\bar{c}}}.  \label{eq:P_ss'}
\end{equation}
In both equations (eqs.~\ref{eq:I(omega)Fourier},~\ref{eq:P_ss'}) it is
assumed that $s\neq s^{\prime }$. The off-diagonal terms $P_{ss^{\prime }}(%
\mathbf{q,}\tau )$ are $P_{LU}(\mathbf{q,}\tau )=P_{UL}^{\ast }(\mathbf{q,}%
\tau )=P_{LL}(\mathbf{q,}\tau )=P_{UU}^{\ast }(\mathbf{q,}\tau )$ and $\bar{c%
}=\alpha ^{2}(0)|\bar{u}_{U0}(\mathbf{q})\bar{u}_{L0}^{\ast }(\mathbf{q}%
)|^{2}$. To perform the integration with respect to $\tau $ on the
right-hand side of Eq.(\ref{eq:I(omega)Fourier}) we use the $\delta $%
-function integral representation $\mathrm{Re}\int_{0}^{\infty }d\tau
e^{ix\tau }=\pi \delta (x)$.

On the next step we take averaging over the phonon bath, i.e. averaging over
$\alpha _{m}$, in the resulting integral Eq.(\ref{eq:I(omega)Fourier}).
Since $\alpha _{m}$ are normally distributed stochastic variables, which
distribution is defined by the density matrix element corresponding to the
excited electronic state
\begin{equation}
\rho _{22}^{(0)}( \alpha _{m}) =[2\pi K(0)]^{-1/2}e^{-\frac{ (\alpha
_{m}-\omega _{st})^{2}}{2K(0)}},  \label{eq:rho_22equilibrium2}
\end{equation}%
their normalized sum, $\alpha (0)=\frac{1}{\mathcal{N}}\sum_{m}\alpha
_{m}\equiv \frac{1}{\mathcal{N}}A$, is also a Gaussian variable. Then the
probability density $\alpha (0)$ is%
\begin{equation}
\rho _{22}\left( \alpha (0)\right) =[2\pi K(0)k_{N}]^{-1/2}e^{ \frac{
(\alpha (0)-\omega _{st})^{2}}{2K(0)k_{N}}}  \label{eq:rho_22(A)N=2}
\end{equation}%
where%
\begin{equation}
k_{N}=(\mathcal{N+}2\sum_{i<j}^{\mathcal{N}}r_{ij})/\mathcal{N}^{2},
\label{eq:k_N}
\end{equation}%
$0\leqslant r_{ij}\leqslant 1$ are the correlation coefficients that are
different from zero when the LFOA vibrations include both the intra- and the
intermolecular ones.

Assumption of the intramolecular nature of the LFOA vibrations, $\alpha _{m}$%
, allows us to conclude that in this case they are statistically
independent, so that the coefficient $k_{N}$ is equal to $k_{N}=1/\mathcal{N}
$. In another extreme case when the LFOA vibrations are intermolecular ones,
correlation coefficients $r_{ij}=1$, and $k_{N}=1$.

The averaging $\langle (...)\rangle $ in Eq.(\ref{eq:I(omega)Fourier}) has
to be done by using the probability density $\rho _{22}\left( \alpha
(0)\right) $ and it reduces to the calculation of the integral $\langle
(...)\rangle =\int_{-\infty }^{\infty }d\alpha (0)\rho _{22}\left( \alpha
(0)\right) (...)$. Using Eqs.(\ref{eq:I(omega)Fourier}), (\ref{eq:P_ss'})
and (\ref{eq:rho_22(A)N=2}), we get
\begin{multline}
I_{p\mathbf{q}_{||}}(\omega )=\frac{4\pi ^{2}\eta _{phot}\hbar \omega _{%
\mathbf{q}_{||}}|u_{\mathbf{q}_{||}}(\mathbf{r)|}^{2}|\tilde{p}_{\mathbf{q}%
}(0)|^{2}}{\sqrt{2\pi k_{N}K(0)T^{2}(\omega )}}  \exp \left\{ -\frac{[(\omega -\Omega _{U})(\omega -%
\Omega _{L})+\omega _{st}T(\omega )]^{2}}{2k_{N}K(0)T^{2}(\omega )}\right\}
\\
\times \frac{(\omega -\Omega _{L})\bar{u}_{U0}+(\omega -\Omega _{U})%
\bar{u}_{L0}}{|T(\omega )|}  
\sum\limits_{s=L,U}[\langle |v_{s}(\mathbf{q)|}^{2}\mathbf{\rangle +\langle }%
v_{s^{\prime }}\mathbf{(\mathbf{q)}}v_{s}^{\ast }\mathbf{(\mathbf{q)\rangle }%
]}(\omega -\Omega _{s^{\prime }})\bar{u}_{s0}
\label{eq:I(omega)1mode2branches2}
\end{multline}%
Here $\tilde{p}_{L\mathbf{q}}(0)=\tilde{p}_{U\mathbf{q}}(0)\equiv \tilde{p}_{%
\mathbf{q}}(0)$, $s\neq s^{\prime }$, $T(\omega )=(\omega -\Omega _{U})|\bar{%
u}_{L0}(\mathbf{q})|^{2}+(\omega -\Omega _{L})|\bar{u}_{U0}(\mathbf{q})|^{2}$
and we consider $\langle v_{s^{\prime }}(\mathbf{q})v_{s}^{\ast }(\mathbf{%
q)\rangle }$ as a real-valued. Note, Eq.(\ref{eq:I(omega)1mode2branches2})
shows that the luminescence spectrum is narrowing with the intramolecular
nature of the LFOA vibrations ($k_{N}=1/\mathcal{N}$) as the number of
molecules increase (the term $k_{N}K(0)=K(0)/\mathcal{N}$ in the exponent),
though such dependence can be compensated at weak exciton-photon coupling as
the Hopfield coefficients become proportional to $Q$ and $\mathcal{N}$, i.e.
$|\bar{u}_{s0}(\mathbf{q})|^{2}\sim Q\sim \mathcal{N}$ (see Eqs.(\ref{eq:Q1}%
), (\ref{eq:u_smv(q)average^2}) and (\ref%
{eq:u_smv(q)average^2-u_sv(q)average^2})). In contrast, for a strong
exciton-photon interaction when $|\bar{u}_{s0}|^{2}$ do not depend on $%
\mathcal{N}$ the spectral narrowing can be observed. The latter depends also
on the strength of correlations between different molecules. The effect is
weakened when the LFOA vibrations include both intra- and intermolecular
contributions, and disappears when the LFOA vibrations are intermolecular
ones ($k_{N}=1$).

The narrowing of the polariton luminescence spectrum by increasing the number
of molecules resembles the exchange (motional) narrowing in the absorption of
molecular aggregates \cite{Knapp84}. The difference lies in the nature of the interaction
responsible for the exchange effects. In molecular aggregate, this is the nearest-neighbour
intermolecular coupling (Coulomb interaction), and in the case of polariton luminescence,
this is a retarded interaction. Indeed, the width of a vibronic transition in a single molecule
is due to the LFOA vibrations. The retarded interaction leads to the delocalization
of a polariton and to averaging over the various configurations of the LFOA intramolecular
vibrational subsystems. This motional narrowing effect reduces the the bandwidth of the molecular vibronic
transitions. It is worthy to note that our model is more general, and
unlike Ref.\cite{Knapp84},
is not limited by static inhomogeneous broadening.

\begin{figure}
\centering
\includegraphics[scale=0.40]{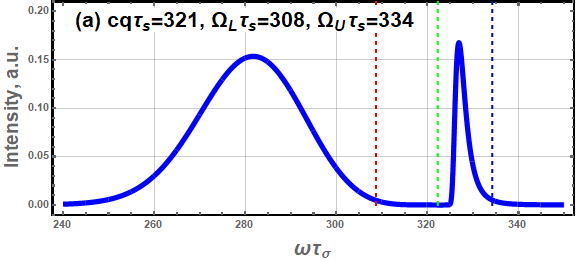} %
\includegraphics[scale=0.40]{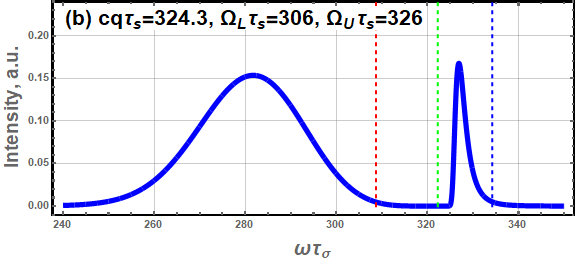}
\caption{ {\small Luminescence spectra of molecules in a single-mode
cavity in the case of the intermolecular nature of the LFOA vibrations (}$%
k_{N}=1${\small ) for $Q\tau _{\sigma }=2$, $K(0)%
\tau _{\sigma }^{2}=120$ and a) $cq\tau _{%
\sigma }=321$, $X_{0m}^{2}<<1$ (a single vibronic transition); b) $cq%
\tau _{\sigma }=324$, $X_{0m}^{2}=0.49$ (involving various vibronic
transitions). The dashed lines correspond to the frequency of the lower
polariton branch (red), upper polariton branch (blue) and to the "zero"
frequency (green).}}
\label{fig:Fano}
\end{figure}

\begin{figure}
\centering
\includegraphics[scale=0.40]{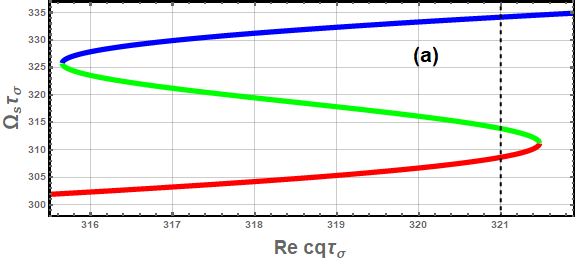} %
\includegraphics[scale=0.40]{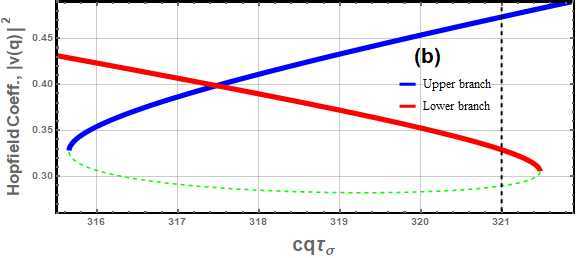} %
\includegraphics[scale=0.40]{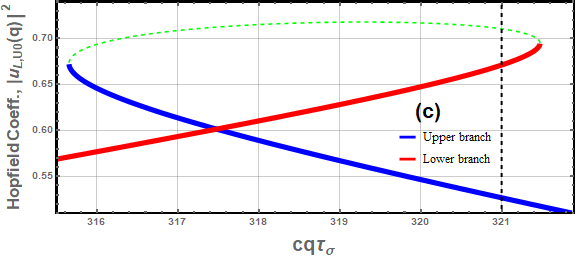}
\caption{{\small The EP dispersion curves (a) plotted for a single
vibronic transition ($X_{0m}^{2}<<1$) at $Q\tau _{\sigma }=2$%
, $K(0)\tau _{\sigma }^{2}=120$; (b) The Hopfield
coefficient $\left\vert v(\mathbf{q})\right\vert^{2}$ and (c) the Hopfield
coefficients $\left\vert \bar{u}_{L0}(\mathbf{q})\right\vert^{2}$ and $%
\left\vert \bar{u}_{U0}(\mathbf{q})\right\vert^{2}$. The blue lines
correspond to the upper polariton branch, and the red lines - to the lower
polariton branch. The vertical black dashed line corresponds to the cavity
mode wavevector chosen for the plot in Fig.~\ref{fig:Fano} a.}}
\label{fig:Disp+Hopfield}
\end{figure}

Our non-Markovian theory enables us to consider both small and large
splitting between the upper and the lower polariton branches compared to the
spectral (homogeneous or inhomogeneous) broadening. The luminescence spectra
of molecules, Eq.(\ref{eq:I(omega)1mode2branches2}), in the case of the
intermolecular nature of the LFOA vibrations, placed in a single-mode cavity
at the splitting having the same order of magnitude as the inhomogeneous
broadening ($K(0)\tau _{\sigma }^{2}=120$) of the molecular spectra, are
presented in Fig.\ref{fig:Fano}. The plots in Fig.\ref{fig:Disp+Hopfield}
show behaviour of the $L$ and $U$ polariton dispersion branches (a); the
Hopfield coefficients $|v_{L,U}(\mathbf{q})|^{2}$ describing the photon
contributions (b); and the coefficients $|\bar{u}_{L,U0}(\mathbf{q})|^{2}$
describing the exciton contributions (c) as functions of the wave number.

It is believed that the polariton states are not formed, when the molecular
resonances broadening is close to, or exceeds the splitting between the
upper and the lower polariton branches. In contrast, for these conditions
our non-Markovian theory demonstrates the main characteristic features of
the Fano resonance \cite{Fano1961PhysRev}. Remind, that the Fano resonance
is a widespread phenomenon associated with the characteristic
\textquotedblleft zero\textquotedblright\ frequency, at which the spectrum
is zero, $I_{p\mathbf{q}_{||}}(\omega _{zero})=0$, and a peculiar asymmetric
and ultra-sharp line shape. Such specific spectral features has found
applications in a large variety of prominent optical devices \cite%
{Joe2006PhysScr,Kivshar2017NatPhot}. One can easily see presence of the zero
frequency from Eq.(\ref{eq:I(omega)1mode2branches2}), it becomes zero when
the term $(\omega -\Omega _{L})\bar{u}_{U0}+(\omega -\Omega _{U})\bar{u}%
_{L0} $ in the right-hand side of Eq.(\ref{eq:I(omega)1mode2branches2}) is
zero. This gives
\begin{equation}
\omega _{zero}=\frac{\bar{u}_{U0}\Omega _{L}+\bar{u}_{L0}\Omega _{U}}{\bar{u}%
_{U0}+\bar{u}_{L0}}.  \label{eq:omega_zero}
\end{equation}%
The \textquotedblleft zero\textquotedblright\ frequency together with the
frequencies of the lower and upper polariton branches are shown in Fig.\ref%
{fig:Fano}.

To conclude the section, we emphasize that the result Eq.(\ref%
{eq:I(omega)1mode2branches2}) predicts the non-Markovian Fano resonance in
the polariton luminescence, which appears because of interference of the
contributions from the upper and the lower polariton branches.

\subsection{Large splitting of the upper and lower polariton branches. Hot
luminescence.}

\label{sec:large_splitting} In the molecular systems based on the organic
dyes\cite{Takazawa10,Fainberg18Advances} the dispersion splitting can reach
the values $\sim 6000-8000$ $cm^{-1}$. At large splitting between the upper
and lower polariton branches the terms with $s=s^{\prime }$ in Eqs.(\ref%
{eq:I(omega)3}) and (\ref{eq:factorization}) produce the main contribution
to the polariton luminescence. For large detuning $\Omega _{s}(\mathbf{q)-}%
\Omega _{s^{\prime }} ( \mathbf{q}) $ ($s\neq s^{\prime }$) the polariton
operators are approximated as follows
\begin{equation}
p_{s\mathbf{q}}(\alpha (0),t-\tau )\approx p_{s\mathbf{q}}(\alpha (0),t)\exp
(id_{s\mathbf{q}}\tau ),  \label{eq:p_sq(alpha,t)}
\end{equation}
so that
\begin{equation}
\sum\limits_{ss^{\prime }}\langle v_{s}(\mathbf{q)}v_{s^{\prime }}^{\ast }(%
\mathbf{q)\rangle \langle }p_{s\mathbf{q}}^{\dag }(t)p_{s^{\prime }\mathbf{q}%
}(t-\tau )\rangle  
\approx \sum\limits_{s}\langle |v_{s}(\mathbf{q)|}^{2}\mathbf{\rangle
\langle }p_{s\mathbf{q}}^{\dag }(t)p_{s\mathbf{q}}(t-\tau )\rangle.
\label{eq:factorization2}
\end{equation}

In the slow modulation limit the average $\mathbf{\langle }p_{s\mathbf{q}%
}^{\dag }(t)p_{s\mathbf{q}}(t-\tau )\rangle $ can be represented as \\ $
\langle p_{s\mathbf{q}}^{\dag }p_{s\mathbf{q}}(\alpha (0),t)\exp (id_{s%
\mathbf{q}}\tau )\rangle $, where $p_{s\mathbf{q}}^{\dag }p_{s\mathbf{q}%
}(\alpha (0),t)=\hat{n}_{s\mathbf{q}}(\alpha (0),t)$ is the polariton
population of the branch $s$, and
\begin{equation}
\mathrm{Re}\int_{0}^{\infty }d\tau \mathbf{\langle }p_{s\mathbf{q}}^{\dag
}(t)p_{s\mathbf{q}}(t-\tau )\rangle \exp (-i\omega \tau )  
=\pi \mathbf{\langle }p_{s\mathbf{q}}^{\dag }p_{s\mathbf{q}}(\alpha
(0),t)\delta (d_{s\mathbf{q}}-\omega ) \rangle.
\label{eq:p^+_sk(t)p_sk(t-tau)}
\end{equation}
 
Since $\alpha (0)=A/\mathcal{N}$ is a stochastic Gaussian variable, one can
average the product $p_{s\mathbf{q}}^{\dag }(t)p_{s\mathbf{q}}(t-\tau )$
over the stochastic process using \ the conditional probability density of
the Gaussian process{\small
\begin{equation}
P\left( \alpha (0),t;\alpha ^{\prime }(0),0\right) =\frac{1}{\sqrt{2\pi
\sigma \left( t\right) /\mathcal{N}}}  
\exp \left\{ -\frac{\left[ (\alpha (0)-\omega _{st})-(\alpha ^{\prime
}(0)-\omega _{st})S\left( t\right) \right] ^{2}}{2\sigma \left( t\right) /%
\mathcal{N}}\right\} .  \label{eq:cond_prob(A)}
\end{equation}
} It is obtained from the conditional probability density distribution of $%
\alpha _{m}$, {\small
\begin{equation}
P\left( \alpha _{m},t;\alpha _{m}^{\prime },0\right) =\frac{1}{\sqrt{2\pi
\sigma \left( t\right) }}  
\exp \left\{ -\frac{\left[ (\alpha _{m}-\omega _{st})-(\alpha _{m}^{\prime
}-\omega _{st})S\left( t\right) \right] ^{2}}{2\sigma \left( t\right) }%
\right\}  \label{eq:cond_prob}
\end{equation} 
} in the case of statistically independent $\alpha _{m}$. In another extreme
case when the LFOA vibrations are intermolecular ones, Eq.(\ref{eq:cond_prob}%
) will be also valid for $\alpha (0)$. To combine these extreme cases we
write{\small
\begin{equation}
P\left( \alpha (0),t;\alpha ^{\prime }(0),0\right) = \frac{1}{\sqrt{2\pi
\tilde{k}_{N}\sigma ( t) }}  
\exp \left\{ -\frac{\left[ (\alpha (0)-\omega _{st})-(\alpha ^{\prime
}(0)-\omega _{st})S ( t ) \right] ^2}{2\tilde{k}_{N}\sigma ( t ) }\right\} .
\label{eq:cond_prob(A)1}
\end{equation}
}The parameter $\tilde{k}_{N}$ equals $1/\mathcal{N}$ for the intramolecular
LFOA vibrations ($\alpha _{m}$ are statistically independent), and $\tilde{k}
_{N}=1$ for the LFOA vibrations having the intermolecular nature. The
variance $\sigma (t)=K(0)\left[ 1-S^{2}\left( t\right) \right] $, and $%
S(t)\equiv K(t)/K(0)$. In the long time limit, the distributions $P\left(
\alpha _{m},t;\alpha _{m}^{\prime },0\right) $ and $P\left( \alpha
(0),t;\alpha ^{\prime }(0),0\right) $ tend to the equilibrium density
matrices of LFOA vibraions in the excited state, Eqs.(\ref%
{eq:rho_22equilibrium2}) and (\ref{eq:rho_22(A)N=2}), respectively. It is
worthy to note that in general $S(t)$ is not necessary an exponential
function. Then the averaging in Eq.(\ref{eq:p^+_sk(t)p_sk(t-tau)}) is
reduced to the calculation of the integral {\small
\begin{multline}
\langle p_{s\mathbf{q}}^{\dag }p_{s\mathbf{q}}(\alpha (0),t)\delta (d_{s%
\mathbf{q}}-\omega )\rangle =n_{s\mathbf{q}}(\alpha ^{\prime }(0),0) \\
\times \int_{-\infty }^{\infty }d\alpha (0)P\left( \alpha (0),t;\alpha
^{\prime }(0),0\right)  \delta \left[ \Omega _{s}(\mathbf{q})-|\bar{u}_{s0}(\mathbf{q}%
)|^{2}\alpha (0)-\omega \right]  \label{eq:int}
\end{multline}%
} with $n_{s\mathbf{q}}(\alpha ^{\prime }(0),0)=\left\langle p_{s\mathbf{q}%
}^{\dag }p_{s\mathbf{q}}(\alpha ^{\prime }(0),0)\right\rangle $. Using Eqs.(%
\ref{eq:I(omega)3}), (\ref{eq:factorization}), (\ref{eq:factorization2}), (%
\ref{eq:p^+_sk(t)p_sk(t-tau)}), (\ref{eq:cond_prob(A)1}) and (\ref{eq:int}),
we get the time-dependent spectrum of the hot (non-equilibrium) luminescence

{\small
\begin{multline}
I_{p\mathbf{q}_{||}}(\omega ,t)=4\pi ^{2}\sum_{s}\eta _{phot}\hbar \omega _{\mathbf{q}%
_{||}}n_{s\mathbf{q}}(\alpha ^{\prime }(0),0)\frac{|u_{\mathbf{q}_{||}}(\mathbf{r)|}^{2}\langle |v_{s}(\mathbf{q)|}%
^{2}\mathbf{\rangle }}{\sqrt{2\pi \tilde{k}_{N}\sigma \left( t\right) |\bar{u%
}_{s0}(\mathbf{q})|^{4}}} \\
\times \exp \Bigg\{-\frac{1}{%
2\tilde{k}_{N}\sigma \left( t\right) |\bar{u}_{s0}(\mathbf{q})|^{4}}   \Big[(\Omega _{s}(\mathbf{q)}-\omega -\omega _{st}|\bar{u}_{s0}(%
\mathbf{q})|^{2})  
-(\alpha ^{\prime }(0)-\omega _{st})|\bar{u}_{s0}(\mathbf{q})|^{2}S\left(
t\right) \Big]^{2}\Bigg\}  \label{eq:I(omega)hot2}
\end{multline}%
} Since $S(0)=1$, the spectrum is centred around the frequency $\Omega _{s}(%
\mathbf{q})-\alpha ^{\prime }(0)$ and is very narrow for $t\rightarrow 0$, $%
I_{p\mathbf{q}_{||}}(\omega ,0)\sim \delta \lbrack \Omega _{s}(\mathbf{q})-
\alpha ^{\prime }(0)-\omega ]$. Remind that Eq.(\ref{eq:I(omega)hot2}) is
derived using the slow modulation limit where the time instant $t$ might be
smaller than $\tau _{\sigma }$, but must be larger than the irreversible
dephasing time of the vibronic transition \cite%
{Fai90OS,Fai93PR,Fai98,Fai00JCCS,Fai03AMPS}. The right-hand side of Eq.(\ref%
{eq:I(omega)hot2}) describes the time-dependent Gaussian spectrum, which
center moves from the initial frequency $\Omega _{s}(\mathbf{q})-\alpha
^{\prime }(0)$ to the frequency $\Omega _{s}(\mathbf{q})-\omega _{st}|\bar{u}%
_{s0}(\mathbf{q})|^{2},$ and broadens over time. For long times $t>>\tau
_{\sigma }$, the spectrum tends to the value{\small
\begin{multline}
I_{p\mathbf{q}_{||}}(\omega ,\infty )=4\pi ^{2}\eta _{phot}\hbar \omega _{%
\mathbf{q}_{||}}|u_{\mathbf{q}_{||}}(\mathbf{r)|}^{2} \\
\times \sum_{s}\frac{\langle |v_{s}(\mathbf{q)|}^{2}\mathbf{\rangle }n_{s%
\mathbf{q}}(\alpha ^{\prime }(0\mathbf{)},0)}{\sqrt{2\pi \tilde{k}_{N}K(0)|%
\bar{u}_{s0}(\mathbf{q})|^{4}}}  
 \exp \left\{ -\frac{\left[ \Omega _{s}(\mathbf{q)}-\omega -\omega
_{st}|\bar{u}_{s0}(\mathbf{q})|^{2}\right] ^{2}}{2\tilde{k}_{N}K(0)|\bar{u}%
_{s0}(\mathbf{q})|^{4}}\right\}  \label{eq:I(omega)equilibrium}
\end{multline}%
} with both the first, $\Omega _{s}(\mathbf{q)}-\omega _{st}|\bar{u}_{s0}(%
\mathbf{q})|^{2}$, and the second central moments, $\tilde{k}_{N}K(0)|\bar{u}%
_{s0}(\mathbf{q})|^{4}$, depending on the exciton contribution to the
polariton. In other words, transition to the polariton degrees of freedom in
our calculations results in replacement of the correlation function, $K(0)$,
and the Stokes shift, $\omega _{st}$, generated by the LFOA vibrations, by
the corresponding quantities multiplied by the factors $|\bar{u}_{s0}(%
\mathbf{q)}|^{4}$ and $|\bar{u}_{s0}(\mathbf{q})|^{2}$ (up to factor $\tilde{%
k}_{N}$), respectively. In contrast, in Ref.\cite{Toppari21JCP}, the authors
were focused on the elastic cavity emission and constrained themselves to a
specific case, when the polariton linewidth is the mean of the cavity and of
the molecular linewidth, although this holds only when the cavity frequency
is in resonance with the electron transition.

\begin{figure}
\centering
\includegraphics[scale=0.40]{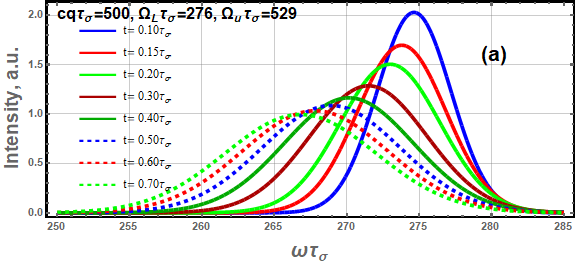} %
\includegraphics[scale=0.40]{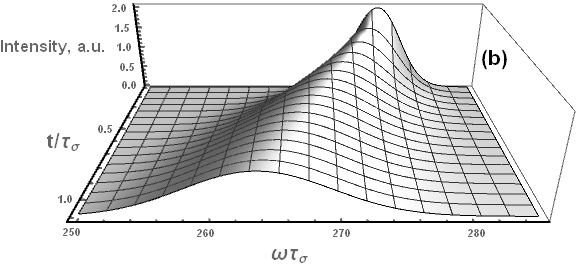}
\caption{ {\small The time-dependent spectrum of molecules in a
single-mode cavity ($cq\tau _{ \sigma }=500$) in the case of
the intermolecular nature of the LFOA vibrations ( $\tilde{k}_{N}=1$), Eq.(
\ref{eq:I(omega)hot2}), plotted as a function of time $t$ and (a) frequency $%
\omega $, and its cross-sections (b) at different instants of time
for $\alpha ^{\prime }(0)=0$. The model parameters are the same as
in Fig.\ref{fig:Dispersion1}.}}
\label{fig:HotLum1}
\end{figure}

The time dependent spectrum of the light emitting polariton system, Eq.(\ref%
{eq:I(omega)hot2}), is shown in Fig.\ref{fig:HotLum1} when the lower
polariton branch is initially excited. Note that since $\bar{u}_{L0}(\mathbf{%
q})\neq \bar{u}_{U0}(\mathbf{q})$, the hot luminescence dynamics differ
quantitatively for the lower and for the upper polariton branches.

It is worthy to note that Eqs.(\ref{eq:I(omega)hot2}) and (\ref%
{eq:I(omega)equilibrium}) do not include further relaxation along the
polariton branches to the value of $\mathbf{q}=(q_{||}= 0,q_{z})$ (see the
comment below Eq.(\ref{eq:d_sk})). This relaxation leads to a further change
in luminescence intensity $I_{p\mathbf{q }_{||}}(\omega )$. The synergy of
the mechanism discussed in this section and the relaxation along the
polariton branches will be considered elsewhere.

\subsection{Averaging with respect to the LFOA vibrations when the slow
modulation limit is not implemented.}

\label{sec:non-Markovian}

In the case of the intramolecular LFOA vibrations, i.e. when the stochastic
variables $\alpha _{m}$ are statistically independent, and the value of $%
\tilde{k}_{N}K(0)\mathcal{=}K(0)/\mathcal{N}$ in Eqs.(\ref{eq:I(omega)hot2})
and (\ref{eq:I(omega)equilibrium}) is much smaller than $K(0)$, which is the
case at large total number of molecules, $\mathcal{N}>>1$, the slow
modulation limit ($\tau _{\sigma }^{2}K(0)/\mathcal{N}>>1$) cannot be taken,
and the semiclassical approximation made in the section "Single mode cavity
and semiclassical approximation", which is based on the frozen nuclear
configuration assumption, ceases to be correct.

In this section we calculate the equilibrium polariton luminescence spectrum
without the constrain of the slow modulation approximation. For the
steady-state conditions the luminescence polariton spectrum is determined by
the Fourier transform of the polariton correlation function $\mathbf{\langle
}p_{s\mathbf{q}}^{\dag }(\tau )p_{s\mathbf{q}}(0)\rangle $ where $p_{s%
\mathbf{q}}^{\dag }(t)$ obeys the hermitian conjugate of Eq.(\ref%
{eq:dp_sk/dt}). Setting $p_{s\mathbf{q}}^{\dag }(t)=\bar{p}_{s\mathbf{q}%
}^{\dag }\exp [i\Omega _{s}(\mathbf{q)}t]$, we obtain
\begin{equation}
\frac{\partial \bar{p}_{s\mathbf{q}}^{\dag }}{\partial t}=-i|\bar{u}_{s0}(%
\mathbf{q})|^{2}\alpha (0 ) \bar{p}_{s\mathbf{q}}^{\dag }  
-i\bar{u}_{s0}^{\ast }(\mathbf{q})\bar{u}_{s^{\prime }0}(\mathbf{q})\alpha (0%
\mathbf{)}\bar{p}_{s^{\prime }\mathbf{q}}^{\dag }e^{i[\Omega _{s^{\prime }}(%
\mathbf{q)}-\Omega _{s}(\mathbf{q})]t},  \label{eq:dp^-_sk0/dt3}
\end{equation}
where $s\neq s^{\prime }$. The expectation value of the polariton operator
obeys the equation
\begin{multline}
\frac{\partial \langle \bar{p}_{s\mathbf{q}}^{\dag }(\alpha (0),t)\rangle }{%
\partial t}=-i|\bar{u}_{s0}(\mathbf{q})|^{2}\alpha (0\mathbf{)}\langle \bar{p%
}_{s\mathbf{q}}^{\dag }(\alpha (0\mathbf{),}t\mathbf{)\rangle } \\
-i\bar{u}_{s^{\prime }0}(\mathbf{q})\bar{u}_{s0}^{\ast }(\mathbf{q})\alpha (0%
\mathbf{)}\langle \bar{p}_{s^{\prime }\mathbf{q}}^{\dag }(\alpha
(0),t)e^{i[\Omega _{s}(\mathbf{q)}-\Omega _{s^{\prime }}(\mathbf{q)]}t}  
+\left. \frac{\partial \langle \bar{p}_{s\mathbf{q}}^{\dag }(\alpha
(0),t)\rangle }{\partial t}\right\vert _{rel}  \label{eq:dp^-_sk0/dt3av}
\end{multline}%
where the last term on the right-hand side describes relaxation due to
interaction of the polaritons with the LFOA vibrations. Using Eqs.(\ref%
{eq:p_sqHF}) and (\ref{eq:a_q,b_mv}), one gets for the derivative in Eq.(\ref%
{eq:dp^-_sk0/dt3av})
\begin{equation}
\left. \frac{\partial \langle \bar{p}\dag _{s\mathbf{q}}(\alpha (0\mathbf{),}%
t\mathbf{)\rangle }}{\partial t}\right\vert _{rel}=\bar{u}_{s0}^{\ast }(%
\mathbf{q})  \left. \frac{\partial }{\partial t}\langle b_{\mathbf{q}0}^{\dag
}(\alpha (0),t)\rangle e^{-i\Omega _{s}(\mathbf{q)}t}\right\vert _{rel}.
\label{eq:dp_sk/dt|rel}
\end{equation}
To proceed we use the property of the Gauss-Markov stochastic process. The
mean value of some function is known to satisfy the differential equation
\begin{equation}
\frac{\partial }{\partial t}\langle b_{\mathbf{q}0}^{\dag }(\alpha _{m}%
,t)\rangle|_{rel}=L_{22}(\alpha _{m})\langle b_{%
\mathbf{q}0}^{\dag }(\alpha _{m},t)\rangle ,  \label{eq:db_q/dt|rel}
\end{equation}%
with the differential operator \cite{Fainberg19JPCC}{\small
\begin{equation}
L_{22}(\alpha _{m})=\tau _{\sigma }^{-1}\left[ 1+\frac{(\alpha _{m}-\omega
_{st})\partial }{\partial (\alpha _{m}-\omega _{st})}+\frac{K(0)\partial ^{2}%
}{\partial (\alpha _{m}-\omega _{st})^{2}}\right] .  \label{eq:L_22(alpha_m)}
\end{equation}%
} Bearing in mind Eqs.(\ref{eq:db_q/dt|rel}) and (\ref{eq:L_22(alpha_m)}) we
rewrite the right-hand side of Eq.(\ref{eq:dp^-_sk0/dt3av}) as
\begin{multline}
\bar{u}_{s0}^{\ast }(\mathbf{q})\left. \frac{\partial }{\partial t}\langle
b_{\mathbf{q}0}^{\dag }(\alpha (0),t)\rangle e^{-i\Omega _{s}(\mathbf{q)}%
t]}\right\vert _{rel} \\
=L_{22}\Big[\bar{u}_{s0}^{\ast }(\mathbf{q)}u_{s^{\prime }0}(\mathbf{q}%
)e^{i(\Omega _{s^{\prime }}(\mathbf{q})-\Omega _{s}(\mathbf{q}))t}\langle
\bar{p}_{s^{\prime }\mathbf{q}}^{\dag }(\alpha (0\mathbf{),}t)\rangle  
+|\bar{u}_{s0}(\mathbf{q})|^{2}\langle \bar{p}_{s\mathbf{q}}^{\dag }(\alpha
(0),t)\rangle \Big]  \label{eq:rel}
\end{multline}%
where the operator $L_{22}(\alpha _{m})$ is now simplified to{\small
\begin{equation}
L_{22}\equiv \tau _{\sigma }^{-1}\left[ 1+\alpha _{2}\frac{\partial }{%
\partial \alpha _{2}}+\tilde{k}_{N}K(0)\frac{\partial ^{2}}{\partial \alpha
_{2}^{2}}\right]  \label{eq:L_22(alpha(0))}
\end{equation}%
} with $\alpha _{2}=\alpha (0)-\omega _{st}$. Here we have used the same
arguments which led us to Eq.(\ref{eq:rho_22(A)N=2}). After the above
simplifications we formulate the set of equation for the functions
\begin{eqnarray}
A(\alpha _{2},t)&=&\exp (-i\Lambda t/2)\langle \bar{p}_{U\mathbf{q}}^{\dag
}(\alpha _{2},t)\rangle 
+\frac{\bar{u}_{L0}}{\bar{u}_{U0}}\exp (i\Lambda t/2\mathbf{)}\langle \bar{p}%
_{L\mathbf{q}}^{\dag }(\alpha _{2},t)\rangle  \label{eq:A(alpha,t)}\\
B(\alpha _{2},t)&=&\exp (-i\Lambda t/2)\langle \bar{p}_{U\mathbf{q}}^{\dag
}(\alpha _{2},t) 
-\frac{\bar{u}_{L0}}{\bar{u}_{U0}}\exp (i\Lambda t/2)\langle \bar{p}_{L%
\mathbf{q}}^{\dag }(\alpha _{2},t)\rangle ,  \label{eq:B(alpha,t)}\end{eqnarray}
where $\Lambda =\Omega _{L}(\mathbf{q})-\Omega _{U}(\mathbf{q})$. The
functions $A(\alpha _{2},t)$ and $B(\alpha _{2},t)$ satisfy the set of
partial-derivative equations
{\small \begin{eqnarray}
\frac{\partial }{\partial t}A(\alpha _{2},t)&=&(|\bar{u}_{U0}|^{2}+|\bar{u}%
_{L0}|^{2})   \lbrack -i\alpha _{2}-i\omega _{st}+L_{22}]A(\alpha _{2},t)  
-i\frac{\Lambda }{2}B(\alpha _{2},t)+A(\alpha _{2},0)\delta (t)\label{eq:dA/dt}\\ \frac{\partial }{\partial t}B(\alpha _{2},t)&=&(|\bar{u}_{U0}|^{2}-|\bar{u}%
_{L0}|^{2})
 (-i\alpha _{2}-i\omega _{st}\mathbf{+}L_{22})A(\alpha _{2},t) 
-i\frac{\Lambda }{2}A(\alpha _{2},t)+B(\alpha _{2},0)\delta (t).
\label{eq:dB/dt}
\end{eqnarray}}
To simplify solution of the equations we also took into account the initial
conditions $A(\alpha _{2},0)$ and $B(\alpha _{2},0)$ by introducing the
terms with the delta-function $\delta (t)$. The inverse transformation from
the functions $A(\alpha _{2},t)$ and $B(\alpha _{2},t)$ to the functions $%
\langle \bar{p}_{s\mathbf{q}}^{\dag }(\alpha _{2},t)\rangle $ has the form
\begin{equation}
\langle \bar{p}_{U\mathbf{q}}^{\dag }(\alpha _{2},t)\rangle =\frac{1}{2}%
e^{i\Lambda t/2}[A(\alpha _{2},t)+B(\alpha _{2},t)],  \label{eq:p_U->A,B}
\end{equation}%
\begin{equation}
\langle \bar{p}_{L\mathbf{q}}^{\dag }(\alpha _{2},t)\rangle =\frac{1}{2}%
\frac{\bar{u}_{U0}}{\bar{u}_{L0}}e^{-i\Lambda t/2}[A(\alpha _{2},t)-B(\alpha
_{2},t)].  \label{eq:p_L->A,B}
\end{equation}
Eqs.(\ref{eq:dA/dt}) and (\ref{eq:dB/dt}) are solved in the Supporting Information.

\subsubsection{Luminescence spectrum calculation}

The equilibrium luminescence spectrum ($t=0$) in a single-mode cavity
according to Eqs.(\ref{eq:I(omega)3}) and (\ref{eq:factorization}) is
written as
\begin{multline}
I_{p\mathbf{q}_{||}}(\omega )=4\pi \eta _{phot}\hbar \omega _{\mathbf{q}%
_{||}}|u_{\mathbf{q}_{||}}(\mathbf{r)|}^{2}\sum_{ss^{\prime }}\langle v_{s}(%
\mathbf{q)}v_{s^{\prime }}^{\ast }(\mathbf{q})\rangle \\
\times \langle p_{s^{\prime }\mathbf{q}}(0)\rangle \mathrm{Re}%
\int_{0}^{\infty }d\tau \mathbf{\langle }\bar{p}_{s\mathbf{q}}^{\prime }\dag
(\tau )\rangle e^{-i(\omega -\Omega _{s}(\mathbf{q}))\tau }
\label{eq:I(omega)steady7}
\end{multline}%
where the polariton operator $\mathbf{\langle }\bar{p}_{s\mathbf{q}}^{\prime
}\dag (\tau )\rangle $ includes averaging over the stochastic process
described by the variable $\alpha _{2}$,
\begin{equation}
\langle \bar{p}_{s\mathbf{q}}^{\prime }\dag (\tau )\rangle =\int d\alpha
_{2}\langle \bar{p}_{s\mathbf{q}}^{\dag }(\alpha _{2},\tau )\rangle
\label{eq:p'_sq(tau)}
\end{equation}%
The latter expression means that the function $\langle \bar{p}_{s\mathbf{q}%
}^{\prime }\dag (\tau )\rangle $ is the Fourier transform of the polariton
operator $\langle \bar{p}_{s\mathbf{q}}^{\dag }(\alpha _{2},\tau )\rangle $
over $\alpha _{2}$, which has to be taken at zero Fourier-conjugate variable
$\varkappa $ (see the Supporting Information). At the same time $\mathbf{\langle }\bar{p}_{s\mathbf{q}%
}^{\prime }\dag (\tau )\rangle $ can be written as the Fourier-transform of $%
\langle \bar{P}_{s\mathbf{q}}^{\prime }\dag (\tilde{\omega}\mathbf{)\rangle }
$
\begin{equation}
\mathbf{\langle }\bar{p}_{s\mathbf{q}}^{\prime }\dag (\tau )\rangle =\frac{1%
}{2\pi }\int_{-\infty }^{\infty }d\tilde{\omega}\exp (i\tilde{\omega}\tau
)\langle \bar{P}_{s\mathbf{q}}^{\prime }\dag (\tilde{\omega}\mathbf{)\rangle
}  \label{eq:p'_sq(tau)2}
\end{equation}%
where $\langle \bar{P}_{s\mathbf{q}}^{\prime }\dag (0\mathbf{,}\tilde{\omega}%
\mathbf{)\rangle =}\int_{-\infty }^{\infty }d\tau \mathbf{\langle }\bar{p}_{s%
\mathbf{q}}^{\prime }\dag (\tau )\rangle \exp (-i\tilde{\omega}\tau )$.
Using Eq.(\ref{eq:p'_sq(tau)2}), the real part of the integral on the
right-hand side of Eq.(\ref{eq:I(omega)steady7}), can be written as%
\begin{equation}
\mathrm{Re}\int_{0}^{\infty }d\tau \mathbf{\langle }\bar{p}_{s\mathbf{q}%
}^{\prime }\dag (\tau )\rangle \exp (-i\tilde{\omega}\tau ) 
=-\mathrm{Im}\frac{1}{2\pi }P\int_{-\infty }^{\infty }d\tilde{\omega}%
^{\prime }\frac{\mathbf{\langle }\bar{P}_{s\mathbf{q}}^{\prime }\dag (0%
 , \tilde{\omega}^{\prime })\rangle }{\tilde{\omega}^{\prime }-\tilde{%
\omega}}  
+\frac{1}{2}\mathrm{Re} \langle \bar{P}_{s\mathbf{q}}^{\prime }\dag
(0 , \tilde{\omega})\rangle,  \label{eq:Re(int)}
\end{equation}
where $\tilde{\omega}=\omega -\Omega _{s}(\mathbf{q)}$. The direct
calculation of the first term on the right-hand side of Eq.(\ref{eq:Re(int)}%
) is difficult. However, this term\ can be expressed in terms of $\mathrm{Re}%
\mathbf{\langle }\bar{P}_{s\mathbf{q}}^{\prime }\dag (0 , \tilde{%
\omega})\rangle $, using \ the Kramers-Kronig relations \cite%
{Fain69,Muk95,NitzanBook2006} (see the Supporting Information)

\begin{equation}
-\frac{1}{\pi }P\int_{-\infty }^{\infty }d\tilde{\omega}^{\prime }\frac{%
\mathrm{Im}\mathbf{\langle }\bar{P}_{s\mathbf{q}}^{\prime \dag }(0 ,
\tilde{\omega}^{\prime })\rangle }{\tilde{\omega}^{\prime }-\omega }=\mathrm{%
Re}\mathbf{\langle }\bar{P}_{s\mathbf{q}}^{\prime \dag }(0 , \tilde{%
\omega})\rangle  \label{eq:ImP->ReP}
\end{equation}%
Then only the real part of the Fourier transform defines the luminescence
spectrum
\begin{equation}
\mathrm{Re}\langle \bar{P}_{s\mathbf{q}}^{\prime \dag }(\tilde{\omega}%
)\rangle =\mathrm{Re}\int_{0}^{\infty }d\tau e^{-i\tilde{\omega}\tau
}\langle \bar{p}_{s\mathbf{q}}^{\prime }\dag (\tau )\rangle
\label{eq:P'_sq(kappa,omega)}
\end{equation}

According to Eqs.(\ref{eq:p_U->A,B}) and (\ref{eq:p_L->A,B}) the function $%
\langle \bar{P}_{s\mathbf{q}}^{\prime \dag }(\tilde{\omega})\rangle $ is a
linear combination of the functions $g(0,\tilde{\omega})$ and $G(0,\tilde{%
\omega})$, Eqs.(\ref{eq:g(k,omega)}), (\ref{eq:g(k,omega)2}) and (\ref{eq:G1(0,omega)2}) of the Supporting Information, with the frequencies shifted by the factor $\Lambda /2$,
\begin{equation}
\langle \bar{P}_{L,U\mathbf{q}}^{\dag }(\tilde{\omega})\rangle =\frac{1}{2}%
\left(
\begin{array}{c}
\bar{u}_{U0}/\bar{u}_{L0} \\
1%
\end{array}%
\right)   \left[ G(0,\tilde{\omega}\pm \frac{\Lambda }{2})\mp g(0,\tilde{\omega}%
\pm \frac{\Lambda }{2})\right]  \label{eq:P'_L,U(kappa,omega)}
\end{equation} 

Using Eqs.(\ref{eq:G1(0,omega)2}), (\ref{eq:g(k,omega)2}) of the Supporting Information and (\ref{eq:P'_L,U(kappa,omega)}), we obtain 
\begin{multline}
\mathrm{Re}\langle \bar{P}_{s\mathbf{q}}^{\prime \dag }(\omega -\Omega
_{s})\rangle =\langle p_{s\mathbf{q}}\dag (0\mathbf{)\rangle } \\
\times \frac{\bar{u}_{L0}(\omega -\Omega _{s^{\prime }})}{T^{2}(\omega )}[%
\bar{u}_{U0}(\omega -\Omega _{L})+\bar{u}_{L0}(\omega  - \Omega _{U})]
 \mathrm{Re}\frac{\Phi (1,1+\bar{x}_{f};\tilde{k}_{N}K(0)\tau _{\sigma
}^{2})}{\tau _{\sigma }^{-1}\bar{x}_{f}}  \label{eq:P_L}
\end{multline}%
where $s,s^{\prime }=L,U$; $s\neq s^{\prime }$, and {\small
\begin{equation}
\bar{x}_{f}=\tilde{k}_{N}K(0)\tau _{\sigma }^{2}+i\tau _{\sigma }\left[
\frac{(\omega -\Omega _{L})(\omega -\Omega _{U})}{T(\omega )}%
+\omega _{st}\right]  \label{eq:x_fL}
\end{equation}%
} Substituting Eqs.(\ref{eq:P'_sq(kappa,omega)}) and (\ref{eq:P_L}) into Eq.(%
\ref{eq:I(omega)steady7}), we finally get {\small
\begin{multline}
I_{p\mathbf{q}_{||}}(\omega )= 4\pi \eta _{phot}\hbar \omega _{\mathbf{q}_{||}}|u_{\mathbf{q}_{||}}(%
\mathbf{r)|}^{2} \frac{\bar{u}_{U0}(\omega -\Omega _{L})+\bar{u}%
_{L0}(\omega -\Omega _{U})}{T^{2}(\omega )}  \mathrm{Re}\frac{\Phi (1,1+\bar{x}_{f};\tilde{k}_{N}K(0)\tau
_{\sigma }^{2})}{\tau _{\sigma }^{-1}\bar{x}_{f}} \\
\times \sum_{s=L,U}|\langle p_{s\mathbf{q}}(0)\rangle |^{2}\bar{u}%
_{s0}(\omega -\Omega _{s^{\prime }})  \left[ \langle |v_{s}(\mathbf{q})|^{2}\rangle +\langle v_{s}(\mathbf{q%
})v_{s^{\prime }}^{\ast }(\mathbf{q})\rangle \right],
\label{eq:I(omega)steady6a}
\end{multline}%
} where $s\neq s^{\prime }$ and the equally distributed initial population
is assumed, $\tilde{p}_{L\mathbf{q}}(0)=\tilde{p}_{U\mathbf{q}}(0)=\tilde{p}%
_{s\mathbf{q}}(0)$.

\begin{figure}
\centering
\includegraphics[scale=0.40]{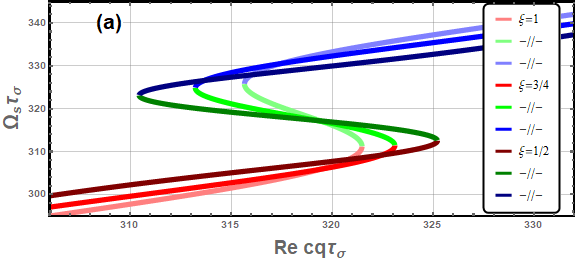} %
\includegraphics[scale=0.40]{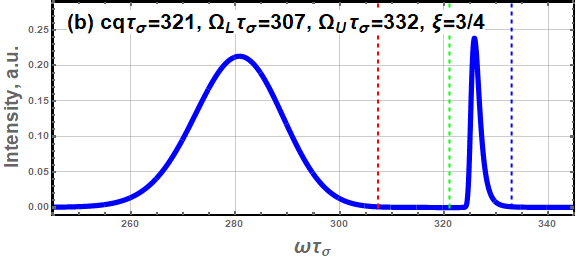} %
\includegraphics[scale=0.40]{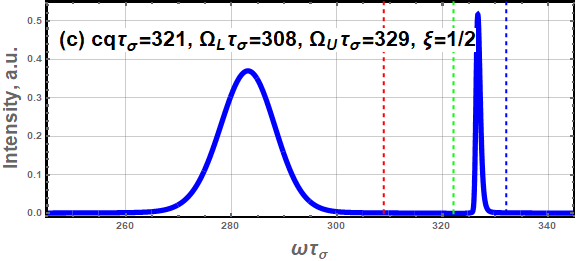} %
\caption{{\small The dispersion curves (a) and the polariton
luminescence spectrum (b, c) of molecules in a single-mode cavity in the case
of the intermolecular nature of the LFOA vibrations ($\tilde{k}_{N}=1$) calculated by
Eq.~(\ref{eq:I(omega)steady6a}) for a single vibronic transition ($%
X_{0m}^2\ll 1$) for various $Q\tau_\sigma= 2\xi$ and $K(0)\tau_\sigma^2=120\xi^2 $, namely
in plot (b) $Q\tau_\sigma=1.5$, $K(0)\tau_\sigma^2=67.5$ and (c) $Q\tau_\sigma=1$
and $K(0)\tau_\sigma^2=30$ (compare with Fig.\ref{fig:Fano} a, where $K(0)\tau_\sigma^2=120$
and $Q\tau_\sigma=2$).}}
\label{fig:VariousQ}
\end{figure}

\begin{figure}
\centering
\includegraphics[scale=0.40]{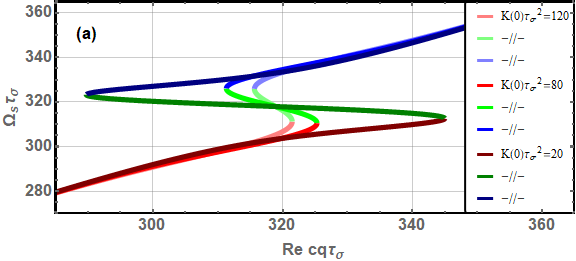} %
\includegraphics[scale=0.40]{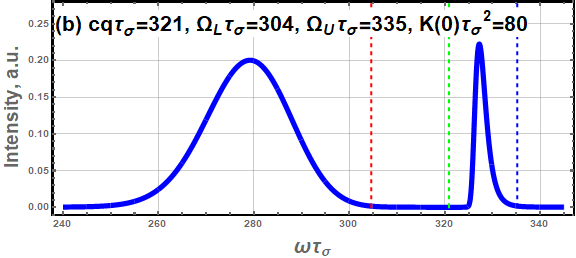} %
\includegraphics[scale=0.40]{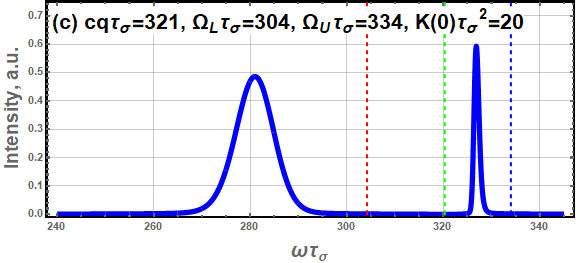}
\caption{\small The dispersion curves (a) and the polariton
luminescence spectrum (b, c) of molecules in a single-mode cavity in the case
of the intermolecular nature of the LFOA vibrations ($\tilde{k}_{N}=1$) calculated by
Eq.~(\ref{eq:I(omega)steady6a}) for a single vibronic transition ($%
X_{0m}^2\ll 1$) for various $K(0)\tau_\sigma^2=80$ (b) and $20$ (c) at $Q\tau_\sigma=2$ (compare with Fig.\ref{fig:Fano} a, where $K(0) \tau_\sigma^2=120$ and $Q\tau_\sigma=2$).
}
\label{fig:VariousK}
\end{figure}

In the slow modulation limit when $\tau _{\sigma }^{2}\tilde{k}_{N}K(0)%
\mathcal{>>}1$, Eq.(\ref{eq:I(omega)steady6a}) transforms into Eq.(\ref%
{eq:I(omega)1mode2branches2}). It is worthy to note that the same factor $%
(\omega  - \Omega _{L})\bar{u}_{U0}+(\omega  - \Omega _{U})%
\bar{u}_{L0}$ is present in both equations which determines the "zero"
frequency, Eq.(\ref{eq:omega_zero}). By this means, the luminescence spectra
described by Eqs.(\ref{eq:I(omega)1mode2branches2}) and (\ref%
{eq:I(omega)steady6a}) have the main characteristic features of the Fano
resonances. Therefore, these equations can be viewed as a generalization of
Fano resonances to the non-Markovian case.
Our numerical results (Fig.~\ref{fig:VariousQ}, \ref{fig:VariousK}) show that the best condition for observing the Fano resonances in polariton luminescence is
the limit of slow modulation, which corresponds to the
inhomogeneous broadening of the molecular spectra, i.e.  the extreme
non-Markovian limit. The condition of the interference between the fluorescent peaks requires also that the splitting between the upper and the lower polariton branches has to be of the same order as the inhomogeneous broadening.

\section{Conclusion}

\label{sec:conclusion}

In our work a theory of equilibrium and non-equilibrium (hot) polariton
luminescence spectra in the polariton basis was developed using the
non-Markovian theory for the description of the polariton interaction with
molecular vibrations. We adopted here a realistic model, according to which
each member of the HFOA vibronic progression in the absorption and
luminescence molecular spectra is broadened due to the presence of the LFOA
vibrations $\{\omega _{\sigma }\}$. At the calculation of the polariton
frequencies and the Hopfield coefficients, we average them over $\alpha _{m}$%
, the electron-vibration coupling related to the LFOA molecular vibrations.
This enabled us to get the dispersion equation that coincides with the
equation for the transverse eigenmodes of the medium.

In spite of averaging of the Hopfield coefficients over the
electron-vibration coupling related to the LFOA molecular vibrations, our
theory demonstrates vibrational relaxation of polaritons after excitation,
since equilibrium positions of molecular vibrations in the ground and
excited states do not coincide (the polaron effect). We have shown that the
frequency shift and the broadening of polariton luminescence spectra
strongly depend on the exciton contribution to the polariton, which itself
is a function of frequency.

Our theory predicts the non-Markovian Fano resonance in the polariton
luminescence due to the interference contributions from the upper and lower
polariton branches, and motional narrowing of the EP luminescence spectrum in the
case of the intramolecular nature of the LFOA vibrations with an increase of
the number of molecules in a single-mode microcavity. In addition, the
theory enables us to consider a non-equilibrium (hot) EP luminescence and
opens a way for its observation in organic-based nanodevices in analogy with
the hot luminescence of molecules and crystals \cite%
{Rebane_Saari78,Freiberg_Saari83}.

The last term on the right-hand side of Eq.(\ref{eq:dp_sk/dt}) describes EP
relaxation along the polariton branches when $\mathbf{q}^{\prime }\neq
\mathbf{q}$. It will be took into account elsewhere to extend our theory to
the multimode cavities, wave guides, and to consider the synergy of the EP
relaxation over wave numbers with the mechanism of the hot EP luminescence
discussed in this paper.

\begin{acknowledgement} The work was supported by the Ministry of Science
\& Technology of Israel (Grant No. 79518) and the Grant RA1900000633 for
cooperation between Ariel University and Holon Institute of Technology.
\end{acknowledgement}


\def\func{\mathrm}
\appendix
\section{Supporting Information}
\subsection{Calculations of the polariton dispersion and the Hopfield coefficients}

The transformation coefficients $u_{smv}$ and $v_{s}(\mathbf{q)}$ in Eq.(\ref%
{eq:p_sHF2}) of the main text and the polariton spectrum $\Omega _{s}$ can
be found by evaluating the commutator $[p_{s},\tilde{H}]/\hbar $. Using Eq.(%
\ref{eq:H(p)2}) and Eq.(\ref{eq:p_sHF2}) together with the total Hamiltonian
(\ref{eq:H^tilde}) we obtain
\begin{multline}
\Omega _{s}p_{s}=\Omega _{s}\left[ \sum_{\mathbf{q}}v_{s}(\mathbf{q)}a_{%
\mathbf{q}}+\sum\limits_{mv}u_{smv}b_{mv}\right]    \\
=\sum_{\mathbf{q}}\left[ \omega _{\mathbf{q}}v_{s}(\mathbf{q)}+\frac{ig}{%
\sqrt{\omega _{\mathbf{q}}}}\sum\limits_{mv}\frac{X_{0m}^{v}}{\sqrt{v!}}e^{-%
\frac{X_{0m}^{2}}{2}+i\mathbf{q\cdot r}_{m}}u_{smv}(\mathbf{q})\right] a_{%
\mathbf{q}}    \\
+\sum\limits_{mv}\left[ u_{smv}(\omega _{2v}+\frac{\omega _{st}}{2}-\alpha
_{m})-i\frac{X_{0m}^{v}}{\sqrt{v!}}\sum_{\mathbf{q}}\frac{g}{\sqrt{\omega _{%
\mathbf{q}}}}e^{-\frac{X_{0m}^{2}}{2}-i\mathbf{q\cdot r}_{m}}v_{s}(\mathbf{q)%
}\right] b_{mv}  \label{eq:Dispersion3}
\end{multline}%
Comparing the coefficients of $a_{\mathbf{q}}$ and $b_{mv}$, we find%
\begin{equation}
(\Omega _{s}-\omega _{\mathbf{q}})v_{s}(\mathbf{q})=\frac{ig}{\sqrt{\omega _{%
\mathbf{q}}}}\sum\limits_{mv}e^{-\frac{X_{0m}^{2}}{2}+i\mathbf{q\cdot r}_{m}}%
\frac{X_{0m}^{v}}{\sqrt{v!}}u_{smv}  \label{eq:v_s(q)3}
\end{equation}%
and%
\begin{equation}
u_{smv}=\frac{-i}{\Omega _{s}\mathbf{-(}\omega _{2v}+\omega _{st}/2)+\alpha
_{m}} \sum_{\mathbf{q}}\frac{g}{\sqrt{\omega _{\mathbf{q}}}}v_{s}(\mathbf{q)}%
e^{-\frac{X_{0m}^{2}}{2}-i\mathbf{q\cdot r}_{m}}\frac{X_{0m}^{v}}{\sqrt{v!%
}}  \label{eq:u_s(q)}
\end{equation}%
Introducing a small decay, the term $1/[\Omega _{s}-(\omega _{2v}+\omega
_{st}/2)+\alpha _{m}]$ on the right-hand side of Eq.(\ref{eq:u_s(q)}) can be
written as%
\begin{equation}
\lim_{\gamma \rightarrow 0}\frac{1}{\Omega _{s}-(\omega _{2v}+\omega
_{st}/2)+\alpha _{m}+i\gamma } =\zeta \lbrack \Omega _{s}\mathbf{-(}\omega
_{2v}+\omega _{st}/2)+\alpha _{m}]  \label{eq:ksi}
\end{equation}%
where $\zeta \lbrack \omega ]=\frac{P}{\omega }-i\pi \delta \lbrack \omega ]$%
, and $P$ stands for the principal value. Substituting Eq.(\ref{eq:ksi})
into Eq.(\ref{eq:u_s(q)}), we get Eq.(\ref{eq:u_s(q)1}) of the main text.

\subsection{The whys and wherefores of taking LFOA vibrations into account on
calculations of the polariton dispersion and the Hopfield coefficients.}

It is worthy to note that in the absense of the LFOA vibrations, the number
of different polariton branches $s$ is not limited by $s=U,L$. In presence
of the HFOA vibrations, instead of two polariton branches one can define a
family of dispersion curves related either to the upper, $U$, or the lower, $%
L$, families. Moreover, the $L$ family branches  can cross each other at $%
X_{0}>1$ (see the figure below). The reason of this intersection can be
explained as follows. At $X_{0}>1$, the vibronic transition $0\rightarrow 1$
and possibly the transition $0\rightarrow 2$ become stronger (the
corresponding absorption intensity peaks increase) than the vibronic
transition $0\rightarrow 0$. In this case, the avoided crossing gap of the $L
$ and $U$ branches corresponding to the transition $0\rightarrow 0$ is
smaller than that for the transitions $0\rightarrow 1$ and $0\rightarrow 2$.
This eventually results in the light-induced intersection of different
branches in the family of lower branches. This means that some of the
polariton frequencies $\Omega _{s}(\mathbf{q})$ can be very close to each
other.

For the HFOA vibration mode the cases $X_{0}<1$ and $X_{0}>1$ correspond to
the limits of weak and strong electron-vibrational coupling, respectively.
Therefore, the approaches based solely on the calculations of the electron
system can be inappropriate for $X_{0}>1$ (see figure below). The lower
family dispersion curves crossings at $X_{0}>1$ lead to the light-induced
conical intersection in the polariton energy manifold, and the topological
effects become important. However, the broadening caused by the presence of
the LFOA vibrations, in our approach, effectively merges the different
branches of the lower dispersion curves family and thus pushes the problem
aside.

\begin{figure}
\centering
\includegraphics[scale=0.45]{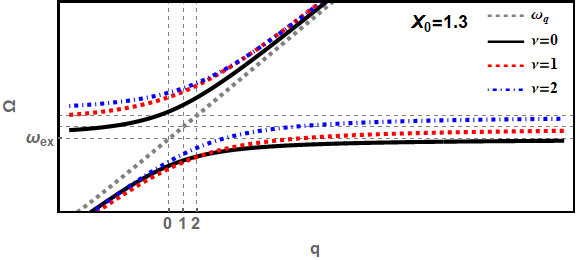}
\caption*{\small
The upper and lower families of the polariton dispersion curves calculated for various vibronic levels at $X_0=1.3$.}
\end{figure}

\subsection{Solution of Eqs.(\ref{eq:dp^-_sk0/dt3av}) of the main text for large detuning $\Omega _{U}(\mathbf{q} )-\Omega _{L}(\mathbf{q}).$}

Introducing the Fourier-transform%
\begin{eqnarray}
\Psi _{s}(\varkappa ,t) &=&\int_{-\infty }^{\infty }d\alpha _{2}\exp
(-i\varkappa \alpha _{2})\langle \bar{p}_{s\mathbf{q}}^{\prime }\dag (\alpha
_{2},t\mathbf{)\rangle },  \label{eq:Psi_s} \\
\langle \bar{p}_{s\mathbf{q}}^{\prime }\dag (\alpha _{2},t\mathbf{%
)\rangle } &=&\frac{1}{2\pi }\int_{-\infty }^{\infty }\Psi _{s}(\varkappa
,t)\exp (i\varkappa \alpha _{2})d\varkappa   \notag
\end{eqnarray}%
and applying it to Eqs.(\ref{eq:dp^-_sk0/dt3av}) of the main text, bearing
in mind Eqs.(\ref{eq:dp_sk/dt|rel}) and (\ref{eq:rel}) we get%
\begin{multline}
\frac{\partial \Psi _{s}(\varkappa ,t)}{\partial t} =\left[-i\omega _{st}+(1%
 - \tau _{\sigma }^{-1}\varkappa )\frac{\partial }{\partial \varkappa
} -\tau _{\sigma }^{-1}\tilde{k}_{N}K(0)\varkappa ^{2}\right]\\\times \left\{|\bar{u}%
_{s0}(\mathbf{q})|^{2}\Psi _{s}(\varkappa ,t)
 +\bar{u}_{s0}^{\ast }(\mathbf{q)}u_{s^{\prime }0}(\mathbf{q})\Psi
_{s^{\prime }}(\varkappa ,t)e^{-i\big(\Omega _{s}(\mathbf{q})-\Omega
_{s^{\prime }}(\mathbf{q})\big)t}\right\}.  \label{eq:Psi_s2}
\end{multline}
Here $s\neq s^{\prime }$. For large detunings $\Omega _{s}(\mathbf{q}%
)-\Omega _{s^{\prime }}(\mathbf{q})$ the last term on the right-hand side of
Eq.(\ref{eq:Psi_s2}) gives small contribution due to fast oscillations and
can be neglected. Setting
\begin{equation}
\Psi _{s}(\varkappa ,t)=\tilde{\Psi}_{s}(\varkappa ,t)e^{-i|\bar{u}_{s0}(
\mathbf{q})|^{2}\omega _{st}t}  \label{eq:Psi_s^tilda}
\end{equation}%
the Eq.(\ref{eq:Psi_s2}) for large detuning becomes
\begin{equation}
\frac{\partial \bar{\Psi}_{s}(\varkappa ,\tilde{t})}{\partial \tilde{t}}%
\approx -\left[(-1\mathbf{+}\tau _{\sigma }^{-1}\varkappa )\frac{\partial }{%
\partial \varkappa } +\tau _{\sigma }^{-1}K^{\prime }(0)\varkappa
^{2}\right]\bar{\Psi}_{s}(\varkappa ,\tilde{t}),  \label{eq:dPsi_s/dt}
\end{equation}%
which is the one-dimensional version of the equation obtained in Ref.\cite%
{Rautian67}. Here $K^{\prime }(0)=\tilde{k}_{N}K(0)$, $\tilde{t}=|\bar{u}%
_{s0}(\mathbf{q})|^{2}t$ and $\bar{\Psi}_{s}(\varkappa ,\tilde{t})=\tilde{%
\Psi}_{s}(\varkappa ,t).$ Returning back to the function $\tilde{\Psi}%
_{s}(\varkappa ,t)$, we rewrite the solution of Eq.(\ref{eq:dPsi_s/dt}) as
\begin{multline}
\tilde{\Psi}_{s}(\varkappa ,t)=B\exp \Bigg\{-\frac{\tilde{k}_{N}K(0)}{2}\Big[%
\varkappa ^{2}+2\varkappa \tau _{\sigma }\left(1-e^{-|\bar{u}_{s0}(\mathbf{q}%
)|^{2}t/\tau _{\sigma }}\right) \\
+2\tau _{\sigma }^{2}\left(|\bar{u}_{s0}(\mathbf{q})|^{2}t/\tau _{\sigma }-1+e^{-|%
\bar{u}_{s0}(\mathbf{q})|^{2}t/\tau _{\sigma }}\right)\Big]\Bigg\}
\label{eq:Psi_s3}
\end{multline}%
with $\tilde{\Psi}_{s}(0,t)=B\tilde{\Psi}_{s}^{\prime }(0,t)$, where%
\begin{equation}
\tilde{\Psi}_{s}^{\prime }(0,t)=\exp \left\{ -\tilde{k}_{N}K(0)\tau _{\sigma
}^{2}\left[ |\bar{u}_{s0}(\mathbf{q})|^{2}t/\tau _{\sigma }-1+e^{-|\bar{u}%
_{s0}(\mathbf{q})|^{2}t/\tau _{\sigma }}\right] \right\}
\label{eq:Psi_s(0,t)}
\end{equation}%
and $B=\langle p_{s\mathbf{q}}\dag (0\mathbf{)\rangle }$ is a constant. Then
the luminescence spectrum given by Eq.(\ref{eq:I(omega)3}) of the main text
can be written for the large detuning as%
\begin{multline}
I_{p\mathbf{q}_{||}}(\omega )=4\pi \eta _{phot}\hbar \omega _{%
\mathbf{q}_{||}}|u_{\mathbf{q}_{||}}(\mathbf{r)|}^{2}\sum_{s}\langle |v_{s}(\mathbf{%
q)|}^{2}\rangle \langle |p_{s\mathbf{q}}(0)|^{2}\mathbf{\rangle } \\
\times \mathrm{Re}\int_{0}^{\infty }d\tau \tilde{\Psi}_{s}^{\prime }(0,\tau
)\exp [i(\Omega _{s}(\mathbf{q)}-|\bar{u}_{s0}(\mathbf{q})|^{2}\omega
_{st}-\omega )\tau ]  \label{eq:I(omega)steady3}
\end{multline}%
where $\tilde{\Psi}_{s}^{\prime }(0,\tau )$ is the characteristic function.
Calculating the integral on the right-hand side of Eq.(\ref%
{eq:I(omega)steady3}), we finally obtain%
\begin{equation}
I_{p\mathbf{q}_{||}}(\omega )=4\pi \eta _{phot}\sum_{s}\hbar \omega _{%
\mathbf{q}_{||}}|u_{\mathbf{q}_{||}}(\mathbf{r)|}^{2}\langle |v_{s}(\mathbf{%
q)|}^{2}\rangle \langle |p_{s\mathbf{q}}(0\mathbf{)|}^{2}\mathbf{\rangle }%
\mathrm{Re}\left[ \frac{\tau _{\sigma }}{|\bar{u}_{s0}(\mathbf{q})|^{2}x_{f}}%
\Phi (1,1+x_{f};\tilde{k}_{N}K(0)\tau _{\sigma }^{2})\right]
\label{eq:I(omega)steady_general}
\end{equation}%
with%
\begin{equation}
x_{f}=\tilde{k}_{N}K(0)\tau _{\sigma }^{2}+i\tau _{\sigma }\left[ \frac{%
\omega -\Omega _{s}(\mathbf{q)}}{|\bar{u}_{s0}(\mathbf{q})|^{2}}+\omega _{st}%
\right]   \label{eq:x_f}
\end{equation}%
and the confluent hypergeometric function $\Phi (1,1+x_{f};\tilde{k}_{N}K%
 ( 0 ) \tau _{\sigma }^{2})$.

In the slow modulation limit, $\tilde{k}_{N}K(0)\tau _{\sigma }^{2}>>1$, the
function $\tilde{\Psi}_{s}^{\prime }(0,\tau )$ on the right-hand side of Eq.~(\ref{eq:I(omega)steady3}) becomes $\tilde{\Psi}_{s}^{\prime }(0,\tau )\simeq
\exp [-\frac{\tilde{k}_{N}K ( 0 ) |\bar{u}_{s0}(\mathbf{q} 
)|^{4}}{2}\tau ^{2}]$. This leads to the polariton luminescence spectrum
given by Eq.~(\ref{eq:I(omega)equilibrium}) of the main text, where $n_{s%
\mathbf{q}}(\alpha ^{\prime }(0 ) ,0)=\langle |p_{s\mathbf{q}}(0%
 )| ^{2} \rangle $, and the bandwidth of which is $\sim \sqrt{%
\tilde{k}_{N}K ( 0 ) |\bar{u}_{s0}(\mathbf{q})|^{4}}$. For
the fast modulation case, $\tilde{k}_{N}K ( 0  )\tau
_{\sigma }^{2}<<1$, and $\tau $ is large, the function $\tilde{\Psi}%
_{s}^{\prime }(0,\tau )$ behaves like $\tilde{\Psi}_{s}^{\prime }(0,\tau
)\simeq \exp [-\Gamma |\bar{u}_{s0}(\mathbf{q})|^{2}\tau ]$, where $\Gamma =%
\tilde{k}_{N}K(0)\tau _{\sigma }$. This results in the Lorentzian-shape
spectrum%
\begin{equation}
I_{p\mathbf{q}_{||}}(\omega )=4\pi \eta _{phot}\hbar \omega _{%
\mathbf{q}_{||}}|u_{\mathbf{q}_{||}}(\mathbf{r)|}^{2}
\sum_{s}\frac{ \Gamma \langle |v_{s}(\mathbf{%
q)|}^{2}\rangle \langle |p_{s\mathbf{q}}(0 )| ^{2} \rangle  |\bar{u}_{s0}(\mathbf{q})|^{2}}{\Gamma ^{2}|\bar{u}_{s0}(%
\mathbf{q})|^{4}+(\Omega _{s}(\mathbf{q)}-|\bar{u}_{s0}(\mathbf{q}%
)|^{2}\omega _{st}-\omega )^{2}}.  \label{eq:I(omega)equilibrium_fast_mod}
\end{equation}

One can see that in both limiting cases of the slow and the fast modulation,
the bandwidth of the polariton luminescence spectrum is proportional to the
square of the Hopfield coefficient corresponding to the exciton contribution
to the polariton, $|\bar{u}_{s0}(\mathbf{q})|^{2}$.

\subsection{Solution of Eqs.(\ref{eq:dA/dt}) and (\ref{eq:dB/dt}) of the main text}

Eqs.(\ref{eq:dA/dt}) and (\ref{eq:dB/dt}) of the main text can be reduced to the second order
equation for $A(\alpha _{2},t)$ {\small
\begin{multline}
\left[\frac{\partial ^{2}}{\partial t^{2}}+\frac{\Lambda ^{2}}{4} -(-i\alpha
_{2}-i\omega _{st}+L_{22})   \left[ \left(|\bar{u}_{U0}|^{2}+|\bar{u}_{L0}|^{2}\right)\frac{\partial }{
\partial t}-i\left(|\bar{u}_{U0}|^{2}-|\bar{u}_{L0}|^{2}\right)\frac{\Lambda }{2}\right]%
\right] A(\alpha _{2},t)\\
=-i\frac{\Lambda }{2}B(\alpha _{2},0 )\delta (t) +
A(\alpha _{2},0)\delta ^{\prime }(t)  \label{eq:d^2A(alpha,t)/dt^2}
\end{multline}%
} Here $\delta ^{\prime }(t)$ denotes the derivative of the $\delta $-function.

Consider the initial conditions. We assume that initially both polariton
branches are equally excited, i.e. $\tilde{p}_{L\mathbf{q}}(0)=\tilde{p}_{U%
\mathbf{q}}(0)=\tilde{p}_{s\mathbf{q}}(0)$. According to Eqs.(\ref%
{eq:A(alpha,t)}) and (\ref{eq:B(alpha,t)}) of the main text
\begin{equation}
A,B(\alpha _{2},0)=\left(1\pm \frac{\bar{u}_{L0}}{\bar{u}_{U0}}\right)\langle \bar{p}_{s%
\mathbf{q}}^{\dag }(\alpha _{2}\mathbf{,}0\mathbf{)\rangle }
\label{eq:A,B(alpha_2,0)}
\end{equation}%
Therefore from Eq.(\ref{eq:A,B(alpha_2,0)}), and Eqs.~(\ref{eq:Psi_s},~\ref{eq:Psi_s3}) we derive the expression for the initial conditions
\begin{equation}
A,B(\alpha _{2},0)=\left(1\pm \frac{\bar{u}_{L0}}{\bar{u}_{U0}}\right)\frac{\langle p_{s%
\mathbf{q}}^{\dag }(0)\rangle }{\sqrt{2\pi \tilde{k}_{N}K(0)}}e^{-\frac{%
\alpha _{2}^{2}}{2\tilde{k}_{N}K(0)}}.  \label{eq:A,B(alpha,0)}
\end{equation}%
The differential equation (Eq.~\ref{eq:d^2A(alpha,t)/dt^2}) of the main text can be
conveniently solved by using the Fourier-transforms with respect to $t$,
\begin{equation*}
a(\alpha _{2},\tilde{\omega})=\int_{-\infty }^{\infty }dte^{-i\tilde{\omega}%
t}A(\alpha _{2},t),
\end{equation*}%
and $\alpha _{2}$,
\begin{equation*}
G(\varkappa ,\tilde{\omega})=\int_{-\infty }^{\infty }d\alpha
_{2}e^{-i\varkappa \alpha _{2}}a(\alpha _{2},\tilde{\omega})
\end{equation*}%
Then from Eq.(\ref{eq:d^2A(alpha,t)/dt^2}) we derive
\begin{multline}
\frac{\partial }{\partial x}G_{1}(x,\tilde{\omega}) 
=\frac{i\omega _{st}\tau _{\sigma }+\frac{i(\tilde{\omega}^{2}-\frac{\Lambda
^{2}}{4})\tau _{\sigma }}{T_{1}(\tilde{\omega})}+x^{2}\tilde{k} _{N}K(0)\tau
_{\sigma }^{2}-1}{1-x}G_{1}(x,\tilde{\omega}) \\
+\frac{\langle p_{s\mathbf{q}}^{\dag }(0\mathbf{)\rangle }[\frac{\Lambda }{2}%
(1-\frac{\bar{u}_{L0}}{\bar{u}_{U0}}) - \tilde{\omega}(1+\frac{\bar{u}%
_{L0}}{\bar{u}_{U0}})]\tau _{\sigma }}{T_{1}(\tilde{\omega})}e^{-\frac{
\tilde{k}_{N}K(0)\tau _{\sigma }^{2}x^{2}}{2}}  \label{eq:dG1(x)/dx}
\end{multline}%
for the function
\begin{equation}
G_{1}(x,\tilde{\omega})=(1-x)G(\tau _{\sigma }x,\tilde{\omega})
\label{eq:G1}
\end{equation}%
with $x=\varkappa /\tau _{\sigma }$ and
\begin{equation}
T_{1}(\tilde{\omega})=|\bar{u}_{U0}|^{2}(\tilde{\omega}-\frac{\Lambda }{2})+|%
\bar{u}_{L0}|^{2}(\tilde{\omega}+\frac{\Lambda }{2})  \label{eq:T1(omega)}
\end{equation}

Integrating Eq.(\ref{eq:dG1(x)/dx}) and using the representation for the
confluent hypergeometric function \cite{Abr64} ($\Gamma (x)$ is the Gamma
function) {\small
\begin{equation}
\Phi (a,b;z)=\frac{\Gamma (b)}{\Gamma (b-a)\Gamma (a)}\int_{0}^{1}\frac{%
dxe^{zx}x^{a-1}}{(1-x)^{1-b+a}},  \label{eq:Fi}
\end{equation}%
} we finally get for $\varkappa =0$ the result
\begin{multline}
G_{1}(0,\tilde{\omega})=G(0,\tilde{\omega}) 
=\frac{\langle p_{s\mathbf{q}}\dag (0)\rangle }{T_{1}(\tilde{\omega})}\left[
(\tilde{\omega}-\frac{\Lambda }{2})+\frac{\bar{u}_{L0}}{\bar{u}_{U0}}(\tilde{%
\omega}+\frac{\Lambda }{2})\right]   \frac{\tau _{\sigma }}{\bar{x}_{f1}}\Phi (1,1+\bar{x}_{f1};\tilde{k}%
_{N}K(0)\tau _{\sigma }^{2})  \label{eq:G1(0,omega)2}
\end{multline}%
where {\small
\begin{equation}
\bar{x}_{f1}=\tilde{k}_{N}K(0)\tau _{\sigma }^{2}+i\tau _{\sigma }\left[
\frac{(\tilde{\omega}-\frac{\Lambda }{2})(\tilde{\omega}+\frac{\Lambda }{2})%
}{T_{1}(\tilde{\omega})}+\omega _{st}\right] \text{,}
\label{eq:x_f2branches2}
\end{equation}%
} Having found expression for $G(\varkappa ,\tilde{\omega})$ we can find the
final expression for the Fourier transform of the second function $B(\alpha
_{2},t)$,
\begin{equation}
g(\varkappa ,\tilde{\omega})=\iint_{-\infty }^{\infty }d\alpha
_{2}dte^{-i\varkappa \alpha _{2}-i\tilde{\omega}t}B(\alpha _{2},t),
\label{eq:g(k,omega)}
\end{equation}

Note that, as it is pointed out in the main text, only the real part of the
functions $g(\varkappa ,\tilde{\omega} )$ and $G(\varkappa ,\tilde{\omega})$
define the equilibrium luminescence spectrum. Therefore under assumption
that the initial population of the polariton branches $\langle p_{s\mathbf{q}%
}^\dag (0 )\rangle $ are real coefficients and using Eqs.(\ref{eq:dB/dt}) of the main text,(\ref{eq:dG1(x)/dx}), and (\ref{eq:G1}) we derive for the real part of the
function $g(\varkappa ,\tilde{\omega})$
\begin{equation}
\mathrm{Re}g(\varkappa ,\tilde{\omega} )= \frac{(\tilde{\omega}-\frac{%
\Lambda }{2})|\bar{u}_{U0}|^{2}-(\tilde{\omega}+\frac{\Lambda }{2})|\bar{u}%
_{L0}|^{2}}{T_{1}(\tilde{\omega})}  \mathrm{Re}G(\varkappa ,\tilde{\omega})  \label{eq:g(k,omega)2}
\end{equation}
This allows us to calculate the spectrum.

\subsection{Calculation of the integral $\frac{1}{\pi }P\int_{-\infty
}^{\infty }d\tilde{\omega}^{\prime }\frac{\func{Im}\langle \bar{P}_{s%
\mathbf{q}}^{\prime }\dag (0,\tilde{\omega}^{\prime }%
)\rangle }{\tilde{\omega}^{\prime }-\tilde{\omega}}$
using\\ the Kramers-Kronig relations}

The direct calculation of the imaginary part of the integral on the
right-hand side of Eq.(100) from the main text is a difficult task.
However, the integral can be recast in terms of the integral of the real part, $\func{Re} \langle \bar{P}_{s\mathbf{q}}^{\prime }\dag (0 , \tilde{ 
\omega})\rangle $ with the help of the Kramers-Kronig relations\cite%
{Fain69,Muk95,NitzanBook2006}. To do this we replace the
 integration variable in the integral representation of the confluent
hypergeometric function, Eqs.(\ref{eq:Fi}) and (\ref{eq:x_f2branches2}) {\small
\begin{multline}
\frac{\Phi (1,1+\bar{x}_{f1};z)}{\bar{x}_{f1}} =\int_{0}^{1}\frac{dxe^{zx}%
}{(1-x)^{2-b}}   \\
=\int_{0}^{\infty }dt\exp [-z(e^{-t}-1+t)]\exp\left \{-i\tau _{\sigma }\left[
\frac{(\tilde{\omega}-\frac{\Lambda }{2})(\tilde{\omega}+\frac{\Lambda }{2})%
}{T_{1}(\tilde{\omega})}+\omega _{st}\right] t\right\}  \label{eq:Fi(1,1+x;z)}
\end{multline}%
}
Consider, first, the case of large detuning $\Omega _{U}(\mathbf{q})-\Omega
_{L}(\mathbf{q})$. Then for the small difference $|\omega -\Omega
_{U}|<<|\omega -\Omega _{L}|$, the function $\langle \bar{P}_{U\mathbf{q}}^{\prime
\dag }(\omega -\Omega _{U})\rangle $  can be written as
\begin{equation}
\langle \bar{P}_{U\mathbf{q}}^{\prime \dag }(\omega -\Omega _{U})\rangle =%
\frac{\langle p_{s\mathbf{q}}\dag (0 )\rangle \tau _{\sigma }}{|\bar{%
u}_{U0}(\mathbf{q)}|^{2}}
  \frac{\Phi (1,1+\bar{x}_{fU};\tilde{k}_N K(0)\tau _{\sigma }^{2} )}{%
\bar{x}_{fU}},
\label{eq:P_U,LDt}
\end{equation}
where
\begin{equation}
\bar{x}_{fU}=\tilde{k}_N K(0)\tau _{\sigma }^{2}+i\tau _{\sigma }\left[
\frac{(\omega -\Omega _{U})}{|\bar{u}_{U0}(\mathbf{q)}|^{2}}+\omega
_{st}\right]   \label{eq:x_fU}
\end{equation}%
using Eqs.(\ref{eq:P_L}) and (\ref{eq:x_fL}) of the main text. Comparing
Eqs.(\ref{eq:Fi(1,1+x;z)}), (\ref{eq:P_U,LDt}) and (\ref{eq:x_fU}), we get%
\begin{equation}
\langle \bar{P}_{U\mathbf{q}}^{\prime \dag }(\omega -\Omega _{U})\rangle
=\langle p_{s\mathbf{q}}\dag (0 )\rangle
 \int_{0}^{\infty }d\tau \tilde{\Psi}_{U}^{\prime }(0,\tau )\exp \{i%
\left[ \Omega _{U}-|\bar{u}_{U0}(\mathbf{q)}|^{2}\omega _{st}-\omega \right]
\tau \}  \label{eq:P_U,LDt2}
\end{equation}
where the characteristic real function $\tilde{\Psi}_{U}^{\prime }(0,\tau )$, Eq.(\ref{eq:Psi_s(0,t)}), can be considered as a relaxation function \cite%
{Kub62R}.

According to Ref.\cite{Fain69}, the susceptibility of a system is defined by%
\begin{equation}
\chi (\tilde{\omega})=\int_{0}^{\infty }d\tau \varphi (\tau )\exp (i\tilde{%
\omega}\tau )  \label{eq:susc1}
\end{equation}%
where $\varphi (\tau )$ is the response function that is related to the
relaxation function by $\tilde{\Psi}_{U}^{\prime }(0,\tau )=\int_{\tau
}^{\infty }dt\varphi (t)$. Then $\chi (\tilde{\omega})$ becomes%
\begin{equation}
\chi (\tilde{\omega})=1+i\tilde{\omega}\int_{0}^{\infty }d\tau \tilde{\Psi}%
_{U}^{\prime }(0,\tau )\exp (i\tilde{\omega}\tau )  \label{eq:susc2}
\end{equation}%
Substituting Eq.(\ref{eq:P_U,LDt2}) into Eq.(\ref{eq:susc2}), we get for the
real and imaginary parts of the susceptibility
\begin{equation}
\chi ^{\prime }(\tilde{\omega})=1+\tilde{\omega}\func{Im}\frac{\langle \bar{P%
}_{U\mathbf{q}}^{\prime }\dag (0 , \tilde{\omega}-|\bar{u}_{U0}(%
\mathbf{q})|^{2}\omega _{st} )\rangle  }{\langle p_{U\mathbf{q}}\dag
(0 )\rangle  }  \label{eq:Re(susc)1}
\end{equation}%
\begin{equation}
\chi ^{\prime \prime }(\tilde{\omega})=\tilde{\omega}\func{Re}\frac{\langle
\bar{P}_{U\mathbf{q}}^{\prime }\dag (0,\tilde{\omega}-|\bar{u}_{U0}(%
\mathbf{q})|^{2}\omega _{st} )\rangle  }{\langle p_{U\mathbf{q}}\dag
(0 )\rangle  }  \label{eq:Im(susc)1}
\end{equation}%
Using the Kramers-Kronig relations \cite{Fain69,Muk95,NitzanBook2006},%
\begin{equation}
\chi ^{\prime }(\tilde{\omega})=\frac{1}{\pi }P\int_{-\infty }^{\infty }d%
\tilde{\omega}^{\prime }\frac{\chi ^{\prime \prime }(\tilde{\omega}^{\prime
})}{\tilde{\omega}^{\prime }-\tilde{\omega}}  \label{eq:Kramers1}
\end{equation}%
\begin{equation}
\chi ^{\prime \prime }(\tilde{\omega})=-\frac{1}{\pi }P\int_{-\infty
}^{\infty }d\tilde{\omega}^{\prime }\frac{\chi ^{\prime }(\tilde{\omega}%
^{\prime })}{\tilde{\omega}^{\prime }-\tilde{\omega}}  \label{eq:Kramers2}
\end{equation}%
we get after some calculations for $\tilde{\omega}^{\prime }/\tilde{\omega}%
\approx 1$
\begin{equation}
\func{Re}\langle \bar{P}_{U\mathbf{q}}^{\prime }\dag (0,\tilde{%
\omega}-|\bar{u}_{U0}(\mathbf{q})|^{2}\omega _{st}\mathbf{)\rangle }=-\frac{1%
}{\pi }P\int_{-\infty }^{\infty }\frac{\func{Im}\langle \bar{P}_{U\mathbf{q}%
}^{\prime }\dag (0,\tilde{\omega}^{\prime }-|\bar{u}_{U0}(\mathbf{q}%
)|^{2}\omega _{st} )\rangle  }{\tilde{\omega}^{\prime }-\tilde{\omega}%
}d\tilde{\omega}^{\prime }  \label{eq:ReP^+->ImP^+}
\end{equation}

The last relation is valid not only for large detunings but also for a more
general case. Indeed, if $\mathbf{\langle }i\bar{P}_{s\mathbf{q}}^{\prime }(0%
,\tilde{\omega})\rangle $ is analytic in the closed upper-half
plane of the complex variable $\tilde{\omega}$ and vanishes as $1/|\tilde{%
\omega}|$ at $|\tilde{\omega}|\rightarrow \infty $ or faster, the following
relation holds

\begin{equation}
\func{Im}\mathbf{\langle }i\bar{P}_{s\mathbf{q}}^{\prime }(0,\tilde{%
\omega})\rangle =-\frac{1}{\pi }P\int_{-\infty }^{\infty }d\tilde{\omega}%
^{\prime }\frac{\func{Re}\mathbf{\langle }i\bar{P}_{s\mathbf{q}}^{\prime }(0%
 , \tilde{\omega}^{\prime })\rangle }{\tilde{\omega}^{\prime }-\omega
}  \label{eq:ReP->ImP}
\end{equation}%
Eq.(\ref{eq:ReP->ImP}) is equivalent to Eq.(\ref{eq:ImP->ReP}) of the main
text.
\providecommand{\latin}[1]{#1}
\makeatletter
\providecommand{\doi}
  {\begingroup\let\do\@makeother\dospecials
  \catcode`\{=1 \catcode`\}=2 \doi@aux}
\providecommand{\doi@aux}[1]{\endgroup\texttt{#1}}
\makeatother
\providecommand*\mcitethebibliography{\thebibliography}
\csname @ifundefined\endcsname{endmcitethebibliography}
  {\let\endmcitethebibliography\endthebibliography}{}

\end{document}